\documentclass[12pt,preprint]{aastex}
\usepackage{graphicx}
\usepackage{url}
\newcommand{\beq}{\begin{equation}}
\newcommand{\eeq}{\end{equation}}
\newcommand{\beqa}{\begin{eqnarray}}
\newcommand{\eeqa}{\end{eqnarray}}

\def\la{\lower.5ex\hbox{$\; \buildrel < \over \sim \;$}}
\def\ga{\lower.5ex\hbox{$\; \buildrel > \over \sim \;$}}

\begin{document}
\title{Dynamics of the Local Group: the Dwarf Galaxies}
\author{P.~J.~E. Peebles}  
\affil{Joseph Henry Laboratories, Princeton University, Princeton, NJ 08544}
\begin{abstract}
I present a dynamical analysis of the measured redshifts and distances of 64 dwarf galaxies at distances between 50\,kpc and 2.6\,Mpc. These dwarfs are assumed to move as test particles in the gravitational field of 12 massive actors---galaxies and groups of galaxies---under the mixed boundary conditions imposed by cosmology. The model fits most of the measured dwarf distances and redshifts. But more work, perhaps on the gravitational interaction among dwarf galaxies, is required to account for the motions of six galaxies in the NGC\,3109 association and two in the DDO\,210 association. The sample of dwarfs is large enough to constrain the halo mass run in the Milky Way. The evidence points to a sharper break from a nearly flat inner rotation curve than predicted by the NFW profile.\end{abstract}
\maketitle

\section{Introduction}\label{sec:1}

Advances in measurements of distances of the numerous dwarf galaxies in and near the Local Group motivate yet another analysis of Local Group dynamics under the initial condition from cosmology that the galaxy peculiar velocities are small and growing at high redshift. The main result is the fit of model to measured dwarf galaxy distances and redshifts presented in Section~\ref{sec:dzalpha}. The fit is reasonably successful for all but eight dwarfs in the NGC 3109 and DDO\,210 associations. The curious properties of these two associations are discussed in Sections~\ref{sec:tracers} and~\ref{sec:challenges}. Two other unexpected results are that the fit to the data on the numerous nearby dwarfs seems to require Milky Mass larger than other recent measurements, and almost twice the mass of M\,31. We may have the data to check these results by combining analyses of inner data from globular cluster positions and motions with outer data from dwarf galaxies (Sec.\,\ref{sec:masses}). 

The dynamical analysis follows Peebles, Tully, and Shaya (2011) and Peebles and Tully (2013), who used the Numerical Action Method (NAM) of dealing with the mixed boundary conditions required by cosmology. Shaya and Tully (2013) analyzed a larger data base by combining NAM for the massive actors and shooting back in time for orbits of the far more numerous less luminous galaxies. Shooting allows efficient analysis of large samples of galaxies. The challenge is to assess the quality of fit of model to data, because shooting always yields a best possible orbit. Presented here is a full NAM analysis (in the numerical method presented in the Appendix in Peebles, Tully, and Shaya 2011). This allows a useful $\chi^2$ measure of the quality of the model fit to the data. The problem with NAM is that the computation time for a fit to redshifts and distances of $N_m$ gravitationally interacting galaxies scales as $N_m^4$. (This is the order $\sim N_m^3$ matrix inversion that drives all the orbits to a solution at a stationary point of the action, multiplied by $\sim N_m$ to drive the $\sim N_m$ parameters in the solution to a minimum of $\chi^2$.) The simplification taken here is to treat as massive just $N_m=12$ objects. This approaches the number readily accommodated by a desktop computer. The many smaller galaxies, in the final solution $N_t = 64$, are treated as massless test particles, or tracers, for which the NAM computation time to a solution that minimizes $\chi^2$ scales as $N_m^3N_t$ (with a large prefactor). This allows analysis of a considerable number of dwarf galaxies in a modest computational effort.

The model for the massive actors presented in Section~\ref{sec:massiveactors} includes seven groups drawn from the Tully (2014) Local Universe catalog\footnote{Available at the Extragalactic Distance Database, {http://edd.ifa.hawaii.edu} as the catalog ``Local Universe (LU)''} along with the Milky Way (MW), M\,31, and three less luminous but particularly interesting galaxies, the Large Magelanic Cloud (LMC), M\,33, and IC\,10. These twelve actors include $98\%$  of the K-band luminosity in the LU catalog at distances less than 6\,Mpc, and perhaps a similar fraction of the mass. This encourages the assumption that the many nearby low luminosity galaxies that are not too close to the massive groups may be treated as massless tracer particles. Motions of galaxies within the groups have their own story to tell, but the focus in this paper is on the motions of the dwarfs outside massive groups treated as rigid masses. Section~\ref{sec:tracers} describes the selection of $N_t=64$ tracers at distances less than $2.6$\,Mpc drawn from McConnachie's (2012; 2015) catalog of low luminosity galaxies.

The large number of nearby tracers with measured distances and redshifts motivates the introduction in Section~\ref{sec:halo shape} of a single-parameter model for the mass distribution in the assumed rigid and spherical halos of MW and M\,31. The method of computation is reviewed in Section~\ref{sec:computation}. The fits to redshifts, distances, proper motions, and the halo shape are presented in Section~\ref{sec:results}, with maps of orbits in Section~\ref{sec:orbits}. Section~\ref{sec:othermodels} compares the model results with other recent analyses. Section~\ref{sec:concludingremarks} offers considerations of where this NAM approach might go next. 

\section{Data, Parameters, and Models}\label{sec:data}
\subsection{Cosmological Model}
The computation is based on the standard cosmologically flat $\Lambda$CDM theory with Friedman equation, neglecting radiation, 
\beq
{1\over a}{da\over dt} = H_o\left[ {\Omega_m\over a^3} + 1 - \Omega_m\right]^{1/2},\qquad
	H_o = 73\hbox{ km s}^{-1}\hbox{ Mpc}^{-1}, \qquad \Omega_m = 0.263. \label{eq:cosmology}
\eeq
Trials that allow Hubble's constant to float to minimize the $\chi^2$ measure of fit to the data in this computation favor $H\sim 80\hbox{ km s}^{-1}\hbox{ Mpc}^{-1}$. Since that seems to be ruled out I take it as appropriate for a study of local extragalactic dynamics to fix $H_o$ in the range favored by astronomical measures of the distance scale (Freedman, Madore, Scowcroft,  et al. 2012; Riess, Macri, Hoffmann, et al.  2016), rather than the smaller value indicated by the CMB anisotropy. The matter density parameter $\Omega_m$ is taken from the CMB constraint on $\Omega H_o^2$.\bigskip

 \begin{figure}[ht]
\begin{center}
\includegraphics[angle=0,width=4.in]{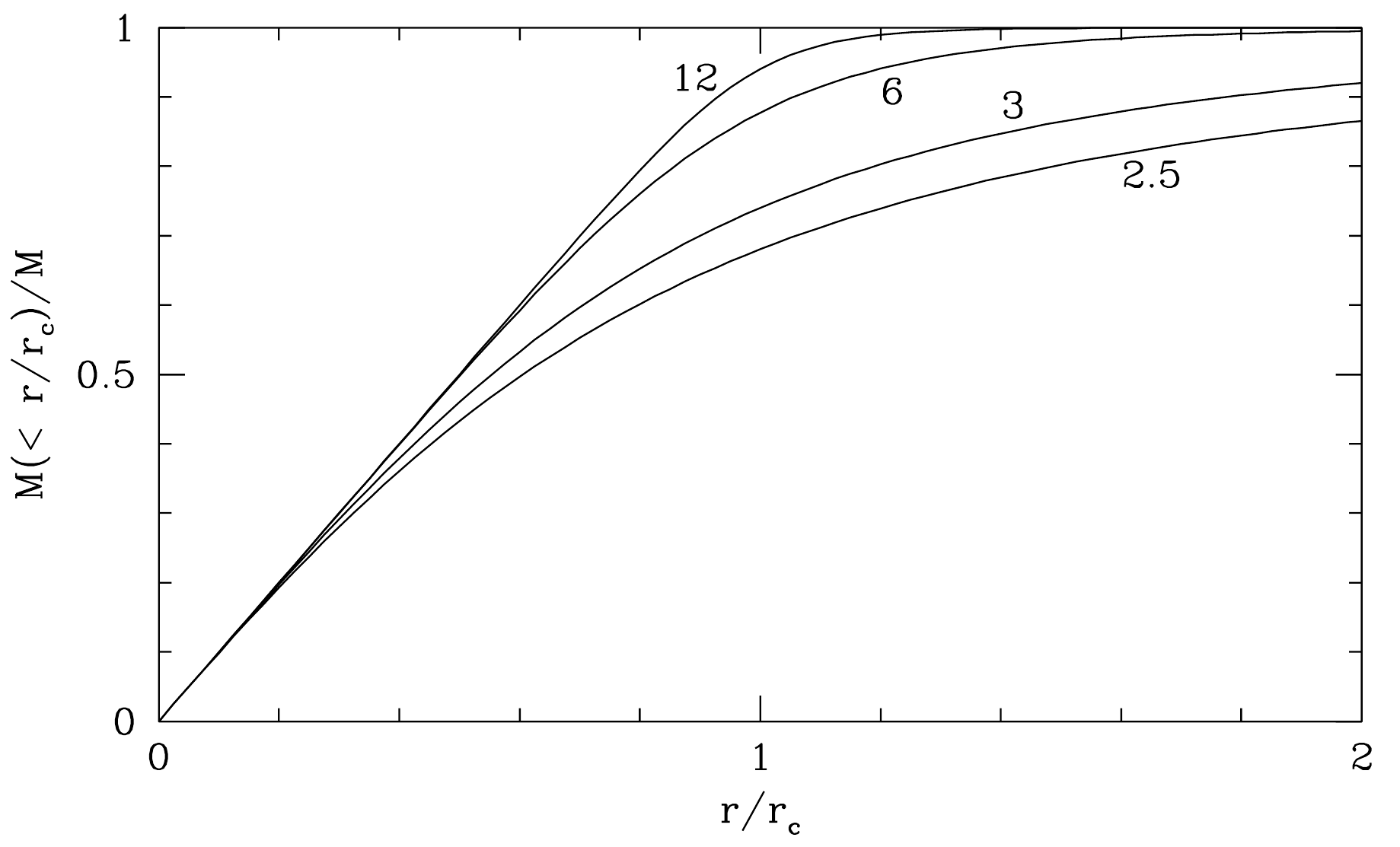} \vspace{-3mm}
\caption{\small Halo Model for the transition from a flat inner rotation curve. The labels are the index $\alpha$ in Equation~(\ref{eq:halomodel}).\label{Fig:halomodel}}
\end{center}
\end{figure}
\vspace{-5mm}

\subsection{Halo Model}\label{sec:halo shape}
Some of the tracers move through the outer parts of the MW or M\,31 halos, and their orbits are influenced by the nature of the radial distribution of the halo mass. I use a one-parameter model in which the mass around each massive actor is assumed to be rigid and spherically symmetric, with density run represented by the gravitational acceleration as a function of physical radius, $r$,  
\beq
g = {v_c^2\over r} \left[1 + {r_c\over r} -\left(1 + \left[r_c\over r\right]^\alpha\right)^{1/\alpha}\right],
\quad r_c={G M\over v_c^2},\quad  \alpha > 1.\label{eq:halomodel}
\eeq
The mass $M$ and circular velocity $v_c$ are assumed to be independent of time in physical units, and $\alpha$ is a dimensionless constant. At $r \ll r_c$ this model approaches the acceleration $g=v_c^2/r$ of a flat rotation curve, and at $r \gg r_c$ it approaches the inverse square law, $g=GM/r^2$. The larger $\alpha$ the sharper the transition between limiting behaviors, as illustrated in Figure~\ref{Fig:halomodel}, and the larger the escape speed at given mass and $v_c$. This is relevant for tracers near MW or M\,31, unimportant for the rest of the tracers. Section~\ref{sec:dzalpha} shows the effect of this halo model on the fit to the data  by comparing results for several choices of $\alpha$. 

I use the following rules for accelerations. The same value of $\alpha$ is used for MW, M\,31, and the seven massive groups. The other three actors are treated as pointlike. The gravitational acceleration exerted by MW on objects other than M\,31 is given by Equation~(\ref{eq:halomodel}) with the chosen value of $\alpha$ and the value of $r_c({\rm MW})$ computed from the MW mass and circular velocity, $v_c({\rm MW})$. The acceleration by M\,31 on objects other than MW  is computed from the M\,31 mass and circular velocity $v_c({\rm M\,31})$. For simplicity I compute the accelerations by MW and M\,31 on each other using  $r_c = [r_c({\rm MW})+r_c({\rm M\,31})]/2$. A realistic treatment of the gravitational interaction of MW and M\,31 would take account of the tidal distortions of the halos, but that is beyond the ambition of this computation. Since MW and M\,31 are well separated this model for their distributed mass has very little effect (as I have checked at $\alpha=6$ by setting $r_c$ close to zero for their interaction). The gravitational accelerations exerted by the seven massive groups are computed using $r_c=100$\,kpc. This is large enough that the tracers have no opportunity to enter close passages with unreasonably  large accelerations, and small enough that the gravitational acceleration of a massive halo on tracers and the other massive actors is very close to the inverse square law. Each interaction conserves momentum.

Conversions of velocities relative to MW to heliocentric redshifts and proper motions use the MW circular velocity $v_c({\rm MW})$ and the solar motion relative to the mean from Sch{\"o}nrich, Binney \& Dehnen  (2010). I allow $v_c({\rm MW})$ to float freely to minimize the $\chi^2$ measure of fit. The M\,31 circular velocity is less important because the catalog distances from M\,31 are far more uncertain than for the closer MW satellites, so I adopt the fixed value $v_c({\rm M31})=230$\,km\,s$^{-1}$ (Corbelli, Lorenzoni, Walterbos, et al.\ 2010). The present  distance to MW is fixed at 8\,kpc. 

 \begin{table}[h]
 \centering
\begin{tabular}{lrrrrrrc}
\multicolumn{8}{c}{Table 1:  Massive Actor Positions, Masses, and Radii} \\
\noalign{\medskip}
\tableline\tableline\noalign{\smallskip}
 actor  & $\ell$\hspace{3mm}  & b\hspace{3mm} & $d^{\rm a}$\hspace{2mm} & 
 	\multicolumn{2}{c}{mass$^{\rm b}$}  & $N_\sigma$ & radius$^{\rm a}$ \\
 \cline{5 - 6}
 & &&&   catalog\hspace{1mm}  &  model\hspace{1mm}  &  & \\
\tableline
   Galaxy &   -\hspace{3mm}   &$    -\hspace{3mm} $&$  -\hspace{3mm}  $&    12.57 &$    28.40 $&$      3.6 $&  --\\
   M31                &   121.17 &$   -21.57 $&     0.77 &    14.24 &$    16.48 $&$      0.6 $&  --\\
   Maff               &   137.06 &$     3.61 $&     3.40 &    55.92 &$    24.92 $&$     -3.1 $&  2.0\\
   M81                &   142.29 &$    40.36 $&     3.59 &    27.88 &$    70.38 $&$      3.5 $&  1.5\\
   Scl                &    69.22 &$   -88.29 $&     3.59 &    23.70 &$    18.20 $&$     -1.0 $&  2.0\\
   Cen                &   311.18 &$    18.87 $&     3.97 &    62.60 &$    35.28 $&$     -2.2 $&  2.5\\
   M94                &   123.60 &$    83.50 $&     4.30 &    24.03 &$    21.05 $&$     -0.5 $&  2.0\\
   N6946              &    96.06 &$    13.14 $&     6.20 &    29.04 &$    17.90 $&$     -1.8 $&  3.0\\
   M101               &   112.25 &$    70.09 $&     7.65 &    95.42 &$   129.18 $&$      1.2 $&  3.0\\
   LMC                &   280.47 &$   -32.89 $&     0.05 &     1.04 &$     2.21 $&$      2.9 $& $-$ \\
   M33                &   133.61 &$   -31.33 $&     0.91 &     0.80 &$     0.84 $&$      0.2 $&  $-$\\
   I10                &   118.96 &$    -3.33 $&     0.79 &     0.36 &$     0.39 $&$      0.3 $&  $-$\\
\tableline
\noalign{\smallskip}
\multicolumn{8}{l}{$^{\rm a}$unit = Mpc \quad $^{\rm b}$unit $=10^{11}M_\odot$}\\
\end{tabular}
\end{table}

\subsection{Mass Model}\label{sec:massiveactors}

The mass in and near the Local Group is assumed to be dominated by the twelve actors listed in Table 1. The seven massive groups of galaxies are meant to represent the mass outside the LG that seems likely to have the largest effect on the orbits of the dwarf galaxies now less than 2.6\,Mpc from MW. The massive groups are named after prominent group members; the names and assignments of members are not sanctioned. I selected the groups from the Tully LU catalog by iteration: center seven spheres on seven of the most luminous LU galaxies at distances less than about 8\,Mpc; compute centers of luminosity of the galaxies in each sphere; recenter the spheres on the centers of luminosity; adjust the sphere radii to the minimum that includes the more luminous neighbors; and iterate to convergence. The last column in Table~1 lists the finally assigned group radii. The angular positions in galactic coordinates and the distances are the luminosity-weighted mean positions of the group members, computed in an orthogonal coordinate system. This is an elaboration and, it is hoped, an improvement of the approach to the mass distribution  outside the Local Group taken in Peebles, Tully, and Shaya (2011) and then Peebles and Tully (2013).

The massive galaxies and halos contain 98\% of the sum of luminosities of LU galaxies at catalog distances less than 3\,Mpc,  98\% of the sum of luminosities at $3 < D < 6$\,Mpc, and  85\% of the sum at $6 < D < 8$\,Mpc. The large fraction included at greater distances is in part the effect of greater incompleteness of lower luminosity galaxies. But if light traces mass the model seems to offer a reasonable description of the distribution of most of the mass within 8\,Mpc from MW, which seems adequate  for the purpose of exploring tracer orbits at distances less than 2.6~Mpc. 

I do not take account of the effect of the tidal field of the mass at greater distance, as was done in the analysis by Shaya and Tully (2013). Consideration of tides in the NAM approach is left for a possible future reanalysis of Local Group dynamics when there is a denser sample of galaxy distances and redshifts beyond 1\,Mpc. 

The nominal catalog masses in Table 1 are the products
\beq
M_{\rm cat} = L_K (d_{\rm model}/d_{\rm cat})^2 M/L_K.\label{eq:MoL}
\eeq
The mass-to-light ratio, $M/L_K$, is the same for all massive actors; its value is allowed to float to minimize $\chi^2$. I term the $M_{\rm cat}$ values catalog, but it is to be understood that these quantities are derived from the catalog luminosities $L_K$ scaled from the LU catalog distances $d_{\rm cat}$ to the model distances $d_{\rm model}$, and that the common factor $M/L_K$ is adjustable. The model mass, apart from MW, is allowed to differ from the nominal catalog value at the penalty
\beq
N_\sigma = \log(M_{\rm model}/M_{\rm cat})/\log(1.3), \label{eq:sdMasses}
\eeq
where $N_\sigma$ is the number of standard deviations from the nominal catalog mass. That is, I allow a 30\% departure from the catalog value at one standard deviation, or a factor of two at a 3-$\sigma$ deviation from the catalog. Since the galaxies MW and M\,31 seem quite similar I assign to the MW mass the penalty
\beq
N_\sigma = \log(M_{\rm MW model}M_{\rm M31 cat}/(M_{\rm MW cat}M_{\rm M31 model}))/\log(1.2), \label{eq:sdMassMW}
\eeq
at $N_\sigma$ standard deviations. The tighter penalty was meant to preserve similar masses of MW and M\,31, but the model with its many constraints prefers a significantly more massive MW. 

The table lists the $\alpha = 6$ model masses  with their numbers of nominal standard deviations from catalog (as discussed further in Sec.~\ref{sec:results}). Four of the twelve masses are 3-$\sigma$ departures from catalog. Since we have little empirical evidence of the relation between an individual galaxy luminosity and its stellar plus dark matter halo mass, and little persuasive theoretical guidance, this  factor of two scatter seems to me intuitively not unacceptable. 

LMC is much less luminous than MW, but my experience suggests that when LMC is treated as massless it is more difficult to find a reasonable LMC orbit. The dynamical significance of the LMC mass is indicated by the model preference for LMC mass twice the catalog value. IC\,10 and M\,33 are even less luminous, but since they are particularly interesting because they have measured proper motions they are included among the twelve massive actors. Since their model masses are close to catalog values I suspect they could have been treated as massless with little effect on the model. (This is difficult to check without repeating the whole computation, because setting the masses of IC\,10 and M\,33 to zero in the present solution seriously disturbs the redshifts and distances of some of the tracers, though it only modestly affects the other massive actors.) 

The orbit of the Small Magellanic Cloud certainly is interesting, but may be intertwined with LMC, as illustrated in Figure~4 in Peebles (2009). A systematic analysis of the SMC history thus seems to be beyond the capabilities of NAM, and this galaxy is not included in the analysis.  

\begin{table}[ht]
 \centering
\begin{tabular}{lccrrrrrr}
\multicolumn{9}{c}{Table 2: Massive Actor Distances and Redshifts} \\
\noalign{\medskip}
\tableline\tableline\noalign{\smallskip}
object & \multicolumn{3}{c} {distance}  & &  \multicolumn{3}{c} {redshift}  & $v_i$ \\
  \cline{2-4} \cline{6-8} \\
\noalign{\vspace{-5mm}}
 &   catalog  &  model & $N_\sigma$  && catalog  & model & $N_\sigma$ & \\
\noalign{\vspace{1mm}}
\tableline
\noalign{\vspace{1mm}}
M31               & $ 0.770\pm 0.040 $ &  0.775 & $   0.1 $ & & $  -301 \pm  10 $&$  -292 $&$  0.9 $ &    63 \\
Maff              & $ 3.397\pm 0.340 $ &  3.938 & $   1.6 $ & & $    12 \pm  30 $&$    16 $&$  0.2 $ &    31 \\
M81               & $ 3.593\pm 0.359 $ &  3.079 & $  -1.4 $ & & $    42 \pm  30 $&$    66 $&$  0.8 $ &    46 \\
Scl               & $ 3.587\pm 0.359 $ &  3.288 & $  -0.8 $ & & $   242 \pm  30 $&$   220 $&$ -0.7 $ &    36 \\
Cen               & $ 3.970\pm 0.397 $ &  3.837 & $  -0.3 $ & & $   518 \pm  30 $&$   515 $&$ -0.1 $ &    41 \\
M94               & $ 4.298\pm 0.430 $ &  4.570 & $   0.6 $ & & $   338 \pm  30 $&$   340 $&$  0.1 $ &    60 \\
N6946             & $ 6.204\pm 0.620 $ &  5.875 & $  -0.5 $ & & $    96 \pm  30 $&$   115 $&$  0.6 $ &    22 \\
M101              & $ 7.647\pm 0.765 $ &  7.261 & $  -0.5 $ & & $   439 \pm  30 $&$   453 $&$  0.5 $ &    21 \\
LMC               & $ 0.050\pm 0.010 $ &  0.061 & $   1.1 $ & & $   271 \pm   5 $&$   266 $&$ -1.0 $ &    40 \\
M33               & $ 0.910\pm 0.050 $ &  0.778 & $  -2.6 $ & & $  -180 \pm   5 $&$  -179 $&$  0.0 $ &    92 \\
I10               & $ 0.794\pm 0.100 $ &  0.968 & $   1.7 $ & & $  -348 \pm   5 $&$  -351 $&$ -0.6 $ &    48 \\
\tableline
\noalign{\smallskip}
\multicolumn{4}{l}{Units: Mpc and km s$^{-1}$} \\
\end{tabular}
\end{table}

Table~2 lists the catalog distances and redshifts of the massive actors. These data are from the LU catalog, but I have adjusted the measurement uncertainties that are treated as standard deviations. M\,31 is assigned redshift uncertainty 10\,km\,s$^{-1}$. The measurement is much more precise, but it seems possible, even likely, that the galaxy of stars is moving relative to the mean of its more massive and extended dark matter halo, at some fraction of the relative motions $\sim 100$\,km\,s$^{-1}$ of stars and gas in different parts of the galaxy. For the same reason, LMC,  M\,33, and IC\,10 are assigned redshift uncertainties 5\,km\,s$^{-1}$, larger than the measurement errors and smaller than for M\,31 because the internal motions are smaller. The redshifts of the seven massive groups are the luminosity-weighted means of the catalog redshifts, and the assigned redshift error is 30\,km\,s$^{-1}$, larger than for M\,31 because the internal motions are larger. 

The distance errors for M\,31, LMC  and  M\,33 are from LU. The low galactic latitude of IC\,10 might make its distance more uncertain. In LU its distance is  $0.79  \pm 0.04$\,Mpc, and in NED the mean is 0.88\,Mpc. I  adopt a larger error flag, $0.79  \pm 0.10$\,Mpc. The  distance error flag for a group is set to 10\% of its distance. This is $\pm 0.8$\,Mpc for the M101 group, an arguably reasonable fraction of its assigned radius of 3\,Mpc. 

Consequences of these intuitive estimates of standard deviations are discussed in Sections~\ref{sec:dzalpha} and~\ref{sec:chisquared}. Here I note that in Table~2 the numbers $N_\sigma$ of standard deviations in the $\alpha=6$ model minus catalog distances are reasonably close to scattering about $\pm 1$, and the $N_\sigma$ for redshifts tend to be unreasonably small. The last column of Table~2 lists the physical velocities $v_i$ of the actors relative to the general expansion of the universe at redshift $1+z=10$. Here again, the $v_i$ tend to be small compared to the assigned standard deviation, 100\,km\,s$^{-1}$ (Sec.~\ref{sec:dzalpha}). A next iteration of this approach might assign less cautious nominal standard deviations. 

\begin{table}[htpb]
 \centering
\begin{tabular}{lccrrrrrr}
\multicolumn{9}{c}{Table 3: Tracer galaxy distances and redshifts} \\
\noalign{\medskip}
\tableline\tableline\noalign{\smallskip}
 & \multicolumn{3}{c} {distance}  & &  \multicolumn{3}{c} {redshift}  & $v_i$ \\
  \cline{2-4} \cline{6-8} \\
\noalign{\vspace{-5mm}}
 &   catalog  &  model & $N_\sigma$  && catalog  & model & $N_\sigma$ & \\
\noalign{\vspace{1mm}}
\tableline
\noalign{\vspace{1mm}}
Bootes (I)        & $ 0.066\pm 0.003 $ &  0.068 & $   0.7 $ & & $    99 \pm   5 $&$   100 $&$  0.2 $ &    83 \\
Draco             & $ 0.076\pm 0.006 $ &  0.076 & $  -0.1 $ & & $  -291 \pm   5 $&$  -291 $&$ -0.1 $ &    52 \\
Ursa Minor        & $ 0.076\pm 0.004 $ &  0.076 & $  -0.1 $ & & $  -246 \pm   5 $&$  -247 $&$ -0.1 $ &    84 \\
Sculptor          & $ 0.086\pm 0.006 $ &  0.085 & $  -0.2 $ & & $   111 \pm   5 $&$   111 $&$ -0.0 $ &    94 \\
Sextans (I)       & $ 0.086\pm 0.004 $ &  0.088 & $   0.4 $ & & $   224 \pm   5 $&$   220 $&$ -0.6 $ &    75 \\
Ursa Major (I)    & $ 0.097\pm 0.005 $ &  0.097 & $  -0.0 $ & & $   -55 \pm   5 $&$   -57 $&$ -0.4 $ &    32 \\
Carina            & $ 0.105\pm 0.006 $ &  0.114 & $   1.4 $ & & $   222 \pm   5 $&$   219 $&$ -0.7 $ &    52 \\
Hercules          & $ 0.132\pm 0.013 $ &  0.159 & $   2.0 $ & & $    45 \pm   5 $&$    45 $&$  0.1 $ &    55 \\
Hydra II          & $ 0.134\pm 0.010 $ &  0.146 & $   1.2 $ & & $   303 \pm   5 $&$   300 $&$ -0.5 $ &    62 \\
Fornax            & $ 0.147\pm 0.013 $ &  0.169 & $   1.7 $ & & $    55 \pm   5 $&$    57 $&$  0.3 $ &    72 \\
Leo IV            & $ 0.154\pm 0.008 $ &  0.144 & $  -1.3 $ & & $   132 \pm   5 $&$   134 $&$  0.3 $ &    67 \\
Canes Venatici II & $ 0.160\pm 0.008 $ &  0.156 & $  -0.5 $ & & $  -128 \pm   5 $&$  -128 $&$  0.0 $ &    68 \\
Leo V             & $ 0.178\pm 0.010 $ &  0.179 & $   0.1 $ & & $   173 \pm   5 $&$   173 $&$  0.1 $ &    77 \\
Canes Venatici (I)& $ 0.218\pm 0.011 $ &  0.223 & $   0.5 $ & & $    30 \pm   5 $&$    33 $&$  0.6 $ &    66 \\
Leo II            & $ 0.233\pm 0.014 $ &  0.242 & $   0.6 $ & & $    78 \pm   5 $&$    74 $&$ -0.8 $ &    70 \\
Leo I             & $ 0.254\pm 0.016 $ &  0.275 & $   1.3 $ & & $   282 \pm   5 $&$   281 $&$ -0.2 $ &    66 \\
Phoenix           & $ 0.415\pm 0.021 $ &  0.438 & $   1.1 $ & & $   -13 \pm   9 $&$   -28 $&$ -1.7 $ &    54 \\
Leo T             & $ 0.417\pm 0.021 $ &  0.369 & $  -2.3 $ & & $    38 \pm   5 $&$    48 $&$  2.1 $ &    95 \\
NGC 6822          & $ 0.459\pm 0.023 $ &  0.442 & $  -0.7 $ & & $   -54 \pm   5 $&$   -61 $&$ -1.3 $ &    73 \\
Andromeda XVI     & $ 0.482\pm 0.038 $ &  0.575 & $   2.4 $ & & $  -367 \pm   5 $&$  -369 $&$ -0.5 $ &    64 \\
Andromeda XXIV    & $ 0.600\pm 0.034 $ &  0.574 & $  -0.8 $ & & $  -128 \pm   5 $&$  -126 $&$  0.3 $ &    90 \\
NGC 185           & $ 0.617\pm 0.031 $ &  0.637 & $   0.6 $ & & $  -203 \pm   5 $&$  -205 $&$ -0.3 $ &    50 \\
Andromeda XV      & $ 0.646\pm 0.059 $ &  0.779 & $   2.3 $ & & $  -323 \pm   5 $&$  -322 $&$  0.2 $ &    47 \\
Andromeda II      & $ 0.652\pm 0.033 $ &  0.598 & $  -1.6 $ & & $  -192 \pm   5 $&$  -189 $&$  0.6 $ &    73 \\
And XXX           & $ 0.658\pm 0.057 $ &  0.679 & $   0.4 $ & & $  -140 \pm   6 $&$  -148 $&$ -1.3 $ &    89 \\
Andromeda X       & $ 0.662\pm 0.033 $ &  0.612 & $  -1.5 $ & & $  -164 \pm   5 $&$  -162 $&$  0.3 $ &    88 \\
NGC 147           & $ 0.676\pm 0.034 $ &  0.676 & $   0.0 $ & & $  -193 \pm   5 $&$  -194 $&$ -0.2 $ &    50 \\
Andromeda XIV     & $ 0.708\pm 0.109 $ &  0.629 & $  -0.7 $ & & $  -480 \pm   5 $&$  -480 $&$ -0.1 $ &    93 \\
Andromeda XXVIII  & $ 0.708\pm 0.124 $ &  0.745 & $   0.3 $ & & $  -326 \pm   5 $&$  -325 $&$  0.1 $ &    58 \\
Andromeda XXIX    & $ 0.731\pm 0.078 $ &  0.711 & $  -0.3 $ & & $  -194 \pm   5 $&$  -194 $&$ -0.1 $ &    70 \\
Andromeda XI      & $ 0.735\pm 0.037 $ &  0.803 & $   1.8 $ & & $  -419 \pm   5 $&$  -419 $&$  0.1 $ &    90 \\
\tableline
\noalign{\smallskip}
\multicolumn{4}{l}{Units: Mpc and km s$^{-1}$} \\
\end{tabular}
\end{table}

\begin{table}[htpb]
 \centering
\begin{tabular}{lccrrrrrr}
\multicolumn{9}{c}{Table 3: continued} \\
\noalign{\medskip}
\tableline\tableline\noalign{\smallskip}
Andromeda XX      & $ 0.735\pm 0.049 $ &  0.831 & $   2.0 $ & & $  -456 \pm   5 $&$  -456 $&$  0.1 $ &    44 \\
IC 1613           & $ 0.755\pm 0.043 $ &  0.667 & $  -2.0 $ & & $  -231 \pm   5 $&$  -224 $&$  1.4 $ &    66 \\
Cetus             & $ 0.755\pm 0.038 $ &  0.726 & $  -0.8 $ & & $   -83 \pm   5 $&$   -82 $&$  0.2 $ &    69 \\
Andromeda XIX     & $ 0.757\pm 0.094 $ &  0.738 & $  -0.2 $ & & $  -111 \pm   5 $&$  -111 $&$ -0.0 $ &    78 \\
Andromeda XXVI    & $ 0.762\pm 0.043 $ &  0.762 & $   0.0 $ & & $  -261 \pm   5 $&$  -261 $&$  0.0 $ &    50 \\
Andromeda VII     & $ 0.762\pm 0.038 $ &  0.803 & $   1.1 $ & & $  -307 \pm   5 $&$  -306 $&$  0.2 $ &    54 \\
Andromeda XXIII   & $ 0.769\pm 0.047 $ &  0.683 & $  -1.8 $ & & $  -237 \pm   5 $&$  -240 $&$ -0.6 $ &    48 \\
LGS 3             & $ 0.769\pm 0.038 $ &  0.793 & $   0.6 $ & & $  -286 \pm   5 $&$  -280 $&$  1.1 $ &    49 \\
Andromeda V       & $ 0.773\pm 0.039 $ &  0.804 & $   0.8 $ & & $  -403 \pm   5 $&$  -402 $&$  0.0 $ &    56 \\
Andromeda VI      & $ 0.783\pm 0.039 $ &  0.752 & $  -0.8 $ & & $  -339 \pm   5 $&$  -340 $&$ -0.2 $ &    49 \\
Leo A             & $ 0.798\pm 0.045 $ &  0.814 & $   0.4 $ & & $    24 \pm   5 $&$    23 $&$ -0.2 $ &    32 \\
Andromeda XXI     & $ 0.826\pm 0.041 $ &  0.781 & $  -1.1 $ & & $  -362 \pm   5 $&$  -362 $&$ -0.0 $ &    52 \\
Andromeda XIII    & $ 0.839\pm 0.042 $ &  0.793 & $  -1.1 $ & & $  -185 \pm   5 $&$  -185 $&$ -0.0 $ &    79 \\
Andromeda XXII    & $ 0.861\pm 0.090 $ &  0.688 & $  -1.9 $ & & $  -129 \pm   5 $&$  -129 $&$  0.1 $ &    69 \\
Andromeda XII     & $ 0.877\pm 0.091 $ &  0.832 & $  -0.5 $ & & $  -558 \pm   5 $&$  -558 $&$  0.0 $ &   100 \\
Tucana            & $ 0.887\pm 0.050 $ &  0.691 & $  -3.9 $ & & $   194 \pm   5 $&$   187 $&$ -1.3 $ &    97 \\
Pegasus dIrr      & $ 0.920\pm 0.046 $ &  0.896 & $  -0.5 $ & & $  -179 \pm   5 $&$  -185 $&$ -1.2 $ &    68 \\
WLM               & $ 0.933\pm 0.047 $ &  0.928 & $  -0.1 $ & & $  -130 \pm   5 $&$  -129 $&$  0.1 $ &    54 \\
DDO210            & $ 1.054\pm 0.105 $ &  1.159 & $   1.0 $ & & $  -108 \pm  30 $&$  -168 $&$ -2.0 $ &    65 \\
Andromeda XVIII   & $ 1.211\pm 0.061 $ &  1.208 & $  -0.0 $ & & $  -332 \pm   5 $&$  -326 $&$  1.1 $ &    91 \\
NGC3109           & $ 1.334\pm 0.133 $ &  1.527 & $   1.5 $ & & $   338 \pm  30 $&$   260 $&$ -2.6 $ &    55 \\
UGC 4879          & $ 1.361\pm 0.068 $ &  1.096 & $  -3.9 $ & & $   -29 \pm   5 $&$   -21 $&$  1.6 $ &    72 \\
KKR 25            & $ 1.923\pm 0.096 $ &  2.125 & $   2.1 $ & & $   -65 \pm  15 $&$  -116 $&$ -3.4 $ &    71 \\
IC 5152           & $ 1.950\pm 0.097 $ &  1.799 & $  -1.6 $ & & $   122 \pm   5 $&$   125 $&$  0.7 $ &    49 \\
KKs3              & $ 2.118\pm 0.106 $ &  2.097 & $  -0.2 $ & & $   316 \pm   7 $&$   316 $&$  0.1 $ &    45 \\
GR 8              & $ 2.178\pm 0.124 $ &  2.295 & $   0.9 $ & & $   213 \pm   5 $&$   212 $&$ -0.3 $ &    55 \\
KKR 3             & $ 2.188\pm 0.124 $ &  2.298 & $   0.9 $ & & $    63 \pm   5 $&$    61 $&$ -0.3 $ &    73 \\
IC 3104           & $ 2.270\pm 0.196 $ &  2.633 & $   1.9 $ & & $   429 \pm   5 $&$   426 $&$ -0.5 $ &    28 \\
UGC 9128          & $ 2.291\pm 0.115 $ &  2.778 & $   4.2 $ & & $   152 \pm   5 $&$   142 $&$ -1.9 $ &    67 \\
IC 4662           & $ 2.443\pm 0.199 $ &  2.372 & $  -0.4 $ & & $   302 \pm   5 $&$   302 $&$  0.1 $ &    51 \\
KKH 98            & $ 2.523\pm 0.126 $ &  2.812 & $   2.3 $ & & $  -136 \pm   5 $&$  -141 $&$ -0.8 $ &    35 \\
UGC 8508          & $ 2.582\pm 0.129 $ &  2.572 & $  -0.1 $ & & $    56 \pm   5 $&$    56 $&$  0.0 $ &    78 \\
KKH 86            & $ 2.582\pm 0.197 $ &  3.302 & $   3.7 $ & & $   287 \pm   5 $&$   281 $&$ -1.1 $ &    69 \\
\tableline\end{tabular}
\end{table}

\subsection{Tracers}\label{sec:tracers}

Table~3 lists distances and redshifts from the McConnachie (2012; 2015) catalog (hereinafter McC), with my assigned standard deviations. The $\alpha=6$ model results are listed with the number $N_\sigma$ of standard deviations of model from catalog. The last column is the physical velocity $v_i$ at redshift $1+z=10$, as in Table~2.

\begin{table}[htpb]
\centering
\begin{tabular}{lccccccr}
\multicolumn{8}{c}{Table 4: Problem galaxies}\\
\tableline\tableline\noalign{\smallskip}
  & \multicolumn{3}{c}{distance$^{\rm a}$} & \  &  \multicolumn{3}{c}{redshift$^{\rm b}$}   \\
   \noalign{\smallskip}
  \cline{2 - 4}  \cline{6 - 8}
 & catalog & model  & $N_\sigma$ && catalog & model  & $N_\sigma$ \\
\tableline
 \noalign{\smallskip}
    Sagittarius dIrr   &    1.067 &    1.584 &$   5.8 $&&$    -78 $&$    -91 $&$  -2.6 $\\
   Aquarius           &    1.072 &    1.395 &$   6.0 $&&$   -137 $&$   -159 $&$  -4.3 $\\
   Antlia B           &    1.294 &    1.838 &$   5.7 $&&$    376 $&$    363 $&$  -2.5 $\\
   NGC 3109           &    1.300 &    1.913 &$   9.4 $&&$    403 $&$    372 $&$  -6.1 $\\
   Antlia             &    1.349 &    1.365 &$   0.2 $&&$    362 $&$    339 $&$  -4.5 $\\
   Sextans B          &    1.426 &    1.985 &$   7.9 $&&$    304 $&$    277 $&$  -5.2 $\\
   Sextans A          &    1.432 &    1.886 &$   6.3 $&&$    324 $&$    304 $&$  -3.9 $\\
   Leo P              &    1.622 &    2.307 &$   4.6 $&&$    264 $&$    255 $&$  -1.6 $\\
\tableline
 \noalign{\smallskip}
\multicolumn{6}{l}{$^{\rm a}$Mpc\quad $^{\rm b}$km\,s$^{-1}$}\\
\end{tabular}
\end{table} 

Three adjustments of the McC data must be discussed. Most important is the treatment of the eight McC galaxies listed in Table~4 with their measured and $\alpha=6$~model redshifts and distances, and the number $N_\sigma$ of nominal standard deviations of model from catalog values. When these eight are included in the solution it more than doubles the $\chi^2$ sum. And it seems significant that these eight galaxies are in two narrow ranges of position and redshift. The last six galaxies in Table~4 are acknowledged  members of the  NGC\,3109\,association (Sand, Spekkens, Crnojevi{\'c}, et al. 2015; McQuinn, Skillman, Dolphin, et al. 2015; Pawlowski and McGaugh 2014; Shaya and Tully 2013; and references therein). The redshifts of these six differ by  140\,km\,s$^{-1}$,  their catalog distances differ by 300\,kpc, and in projected separation they are spread across 800\,kpc. I have lumped them in an association that I assume moves as a single tracer galaxy. The other two problem galaxies, Aquarius and Sagittarius dIrr, are at projected separation $\sim 300$\,kpc, redshift difference 60\,km\,s$^{-1}$, and at the same catalog distance. I do not know a precedent, but their proximity is my excuse for lumping them in the DDO\,210 association (another name for the Aquarius dwarf irregular galaxy). I take the distances and redshifts of the two associations to be the unweighted means of the member distances and redshifts, and assign standard deviations 30\,km\,s$^{-1}$ in redshift and 10\% in distance. These associations treated as tracers with my generous standard deviations fit in the model reasonably well. (A few other points might be noted. In an earlier study with fewer galaxies and a more schematic treatment of the massive actors, Peebles, Tully, and Shaya 2011 were unable to fit Sagittarius dIrr and Aquarius, which I now lump in the DDO\,210 association. This earlier model did fit Sextans\,A and~B. The other four galaxies in the NGC\,3109 association were not in the study.  Table~4 shows the distances and redshifts in the $\alpha=6$ model when the eight problem galaxies are counted in the total $\chi^2$ sum. It may be significant that all the model distances in the associations are too large and all the model redshifts are too small. Adjustment of parameters to minimize the large values of $N_\sigma$ for these galaxies increases discrepancies between model and catalog for other tracers and massive actors. When the eight are replaced by the two associations treated as tracers, the five largest remaining discrepancies in distance or redshift are  about $4\sigma$. This discussion continues in Sections~\ref{sec:chisquared} and~\ref{sec:challenges}.)  

The second adjustment to the data is the removal of McC galaxies close to massive actors. I remove the seven McC galaxies with catalog distances from M\,31 less than 100\,kpc, because the fractional uncertainties in distances from M\,31 are particularly large. I exclude McC galaxies closer to MW than LMC, because their orbits seem likely to be too complicated for NAM. And I remove the McC galaxies that have catalog positions within the seven spheres that define the massive groups. That removes UGCA\,86 at 600 kpc from the actor Maff; NGC\,55, ESO\,294-G, NGC\,300, UKS\,2323-3, and K\,258 at 1.6 to 1.9\,Mpc from the actor Scl, and DDO\,125,  DDO\,99, DDO\,190, NGC\,4163, and DDO\,113 at 1.4 to 1.8\,Mpc from the actor M94. This leaves the 64 tracers at catalog distances less than 2.58\,Mpc listed in Table~3.

The third adjustment is the assignment of nominal standard deviations of the measured distances and redshifts. The art of galaxy distance measurements may be capable of producing distances that are more precise than accurate. Perhaps this is the case for a few McC entries with fractional distance errors less than about 2\%. I take the standard deviation in distance to be the larger of the catalog value or 5\% of the catalog distance. To take account of possible systematic motion of the gas and stars relative to the dominant dark matter, and the systematic error introduced by the schematic nature of the mass model, I set the uncertainties in redshifts of the McC galaxies to the larger of the catalog value or 5\,km\,s$^{-1}$.

\begin{table}[ht]
\centering
\begin{tabular}{lrrrrr}
\multicolumn{6}{c}{Table 5: Proper Motions$^{\rm a}$}\\
\noalign{\medskip}
\tableline\tableline\noalign{\smallskip}
  & \multicolumn{2}{c}{$\mu_\alpha$} & \  &  \multicolumn{2}{c}{$\mu_\delta$}  \\
   \noalign{\smallskip}
  \cline{2 - 3}  \cline{5 - 6}
 & measured\hspace{4mm} & $N_\sigma$  &\ & measured\hspace{4mm} & $N_\sigma$  \\
\tableline
 \noalign{\smallskip}
  M31                &$    0.044 \pm    0.013 $&$   2.3 $&\ &$   -0.032 \pm    0.012 $&$   1.0 $ \\
  LMC                &$    1.891 \pm    0.032 $&$   0.7 $&\ &$    0.226 \pm    0.050 $&$  -0.2 $ \\
  M33                &$    0.023 \pm    0.006 $&$   0.9 $&\ &$    0.002 \pm    0.007 $&$  -3.1 $ \\
  I10                &$   -0.002 \pm    0.008 $&$   1.5 $&\ &$    0.020 \pm    0.008 $&$  -2.1 $ \\
  Leo I              &$   -0.114 \pm    0.029 $&$  -1.9 $&\ &$   -0.126 \pm    0.029 $&$   0.7 $ \\
\tableline
 \noalign{\smallskip}
\multicolumn{6}{l}{$^{\rm a}$milli arc sec y$^{-1}$}\\
\end{tabular}
\end{table}

\subsection{Proper Motions}

Table 5 lists measured proper motions for four massive actors and one tracer.  The proper motion of M\,31 is from van der Marel, Fardal, Besla., et al. (2012a), where I convert their proper velocities to angular velocities using their M31 distance, 770 kpc. For LMC I use the means of values and uncertainties in Kallivayalil, van der Marel,  Anderson, and Alcock (2013) and  van der Marel and  Sahlmann (2016). The reference for M33 is Brunthaler, Reid, Falcke,  Greenhill, and Henkel (2005); for IC\,10 it is Brunthaler, Reid, Falcke, et al. (2007); and for Leo\, I it is Sohn, Besla,  van der Marel,  et al. (2013). I do not attempt to fit the proper motion of Draco because the stated uncertainties seem small compared to the differences among the several measurements (Casetti-Dinescu and Girard 2016, Fig.~11). As for the other parameters, the model results in Table~5 (for $\alpha = 6$) are expressed as the number $N_\sigma$ of standard deviations of model minus measurement.   

\subsection{Measure of Fit}

I treat as standard deviations the estimates of uncertainties in the values of the 12 masses (defined in Eqs.~[\ref{eq:sdMasses}] and~[\ref{eq:sdMassMW}]), the 75 distances and 75 redshifts (listed in Tables 2 and 3), the 10 components of proper motion (Table~4), and the 76 peculiar velocities $v_i$ at $1+z=10$ (listed in the right-hand columns of Tables~2 and~3, and with assigned standard deviation $\sigma_{v_i}=100$\,km\,s$^{-1}$). The measure of fit is the sum over the  248 terms
\beq
\chi^2 = \sum (N_\sigma)^2, \qquad  
N_\sigma = {\hbox{model} - \hbox{catalog}\over\hbox{standard deviation}}.\label{eq:chisquared}
\eeq
(A proper analysis would take account of the three components of each initial peculiar velocity treated as a Gaussian random variable, but that is too fine.) For an assessment of significance one might reduce the count of terms by 75, for the freedom to adjust distances to reduce $\chi^2$, and another three, for the freedom to adjust $M/L_K$, the MW circular velocity $v_c$, and the halo shape index $\alpha$. That would leave expected value  $\chi^2\sim 170$, if the model  and standard deviations were sufficiently accurate. In the best models the values of $\chi^2$ are twice that. The situation is discussed in Section~\ref{sec:chisquared}.

\section{Computation}\label{sec:computation}

The physical assumptions and numerical methods for this NAM analysis are presented in Peebles (2009, 2010), Peebles, Tully, and Shaya (2011), and references therein. The massive actors are assigned constant rigid distributed masses (Sec.\,\ref{sec:halo shape}). The actors and the massless tracers are assumed to move under pure gravity. (The effect of the cosmological constant appears in the expansion parameter $a(t)$ as a function of time.) A more realistic analysis would take account of the merging of mass in the formation of each galaxy. I assume the motion of a model actor or tracer at high redshift usefully approximates the motion of the center of the mass that is gathering together to form the object. This assumption has not been tested in numerical simulations of galaxies in the $\Lambda$CDM cosmology.  But although this cosmology passes demanding tests on the scales of clusters of galaxies and larger, and offers a viable picture of galaxy formation, the picture is not yet very predictive. Thus I attempt to follow a more empirical approach to Local Group dynamics.

The treatment of boundary conditions requires special mention. One can use the measured distances, redshifts, and proper motions of galaxies for a meaningful estimate of their orbits forward and backward in time, as in van der Marel, Besla, Cox, et al. (2012b). The accuracy may be quite limited, however. In particular, the inevitable errors in the measured present conditions cause the peculiar velocity of the orbit computed back in time to diverge. The effect is illustrated in Figure 1 in Peebles and Tully (2013). It can be compared to the situation in linear perturbation theory, where numerical error in the mass density contrast and peculiar velocity field assigned at a given time introduces an artificial decaying component. That decaying component predicts large departures from homogeneity at high redshift, contrary to the growth of structure in the established cosmology.  The variational Numerical Action Method used in this computation yields solutions under the mixed boundary conditions that the peculiar velocities of the particles are increasing with increasing time at high redshift and end up at assigned present positions.

The numerical solutions presented in the next section were found by randomly casting orbits, either placing the position of the galaxy at random at each time step, or placing positions along a linear or spiral function of $a(t)$,  and then adjusting the orbits to a zero derivative of the action by moving the position at each time step in the direction indicated by the first and second derivatives of the action. Most of the reasonably acceptable tracer orbits were found after a few tens of trials; some were found only after a few thousand trials; but $10^5$ trials failed to do much better. A promising solution was then shifted toward lower $\chi^2$ by iterative adjustments of the distances, masses, $v_c$, and $M/L_K$.

The NAM orbits were computed from redshift $z=9$ to the present, a factor of ten expansion, in 500 steps equally spaced in $a(t)$. At a stationary point of the action the orbit is a numerical solution to the usual equation of motion in leapfrog approximation. This numerical solution was checked for accuracy by a conventional leapfrog numerical integration forward in time in 5000 steps in $a(t)$ from $1+z=10$. One can get quite precise initial velocities at $a=0.1$ from the NAM solution because at high redshift the coordinate positions are changing linearly with $a(t)$ to good accuracy. The integration of all the orbits forward in time from the NAM initial positions and velocities usually arrives at the given present positions and velocities within 0.1\,kpc and 0.5\,km\,s$^{-1}$. In particularly curved orbits, such as for LMC, the differences are never greater than 0.5\,kpc and 1.5\,km\,s$^{-1}$. I conclude from this consistency for time steps differing by a factor of ten that numerical errors are negligible.

\section{Results}\label{sec:results}

\subsection{Redshifts, Distances, Proper Motions, Masses, and the MW Halo}\label{sec:dzalpha}

The numbers $N_\sigma$ of nominal standard deviations of the $\alpha=6$ model masses from the  catalog masses are listed in Table~1, the $N_\sigma$ differences of model from catalog redshifts and distances are in Tables~2 and~3, and  the $N_\sigma$ differences of model from catalog proper motions are in Table~4. 

\begin{figure}[ht]
\begin{center}
\includegraphics[angle=0,width=6.5in]{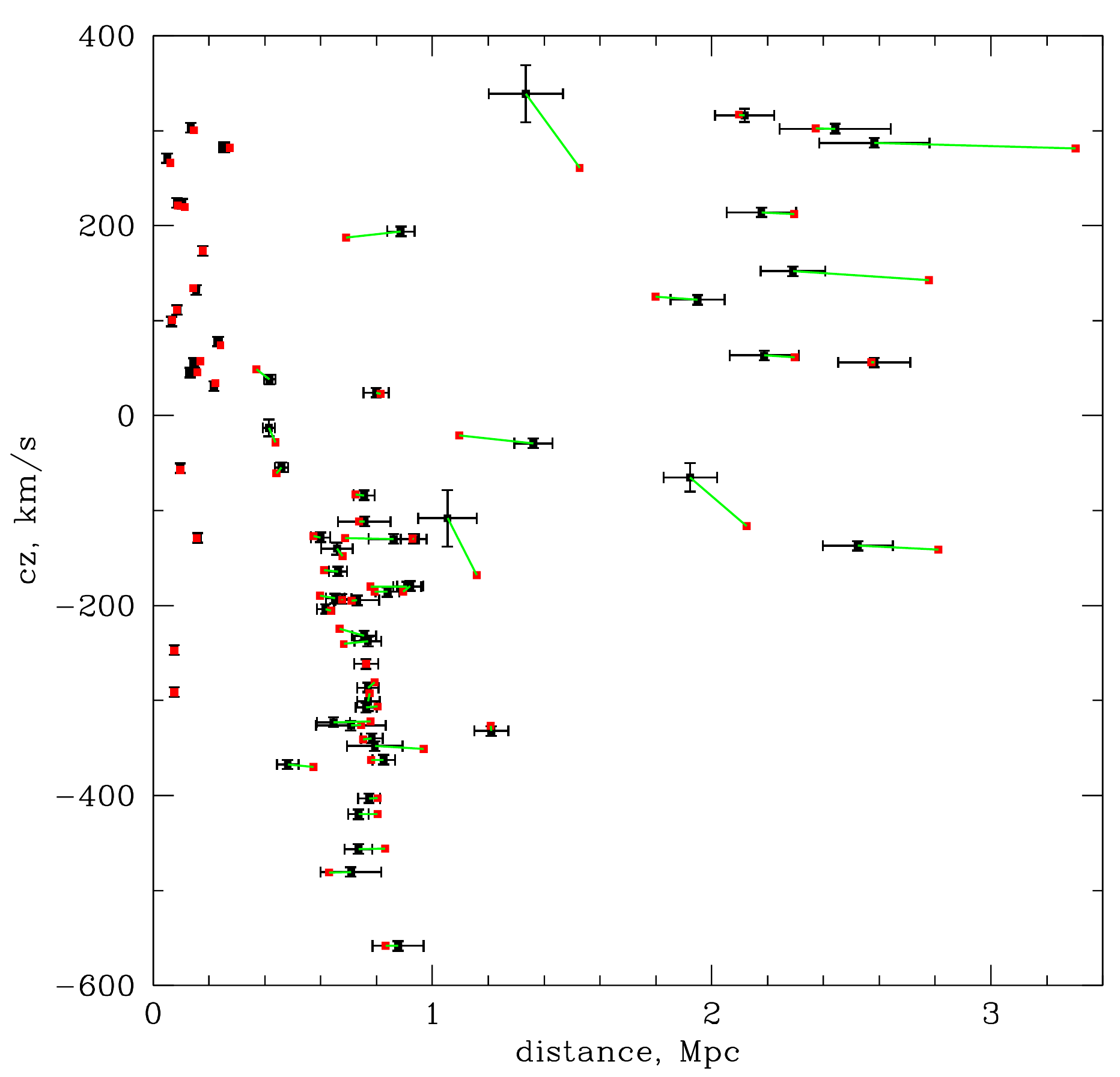} \vspace{-10mm}
\caption{\small Catalog distances and redshifts with error flags are plotted in black, model values in red, and lines connecting model and catalog values in green.\label{Fig:dz}}
\end{center}
\end{figure}
Figure~\ref{Fig:dz} compares catalog and $\alpha=6$ model redshifts and distances for LMC, M\,31, M\,33, IC\,10, and the wealth of data in the McC catalog. The seven massive groups are off scale or close to the right-hand edge of the figure, and clutter is reduced by not showing them. Black squares with error flags mark the catalog values, the red squares are the values in the $\alpha=6$ model, and the green lines connect model and catalog. The figure shows the successful model fits to the scatter of an impressive number of positive and negative redshifts relatively close to MW, at distances less than about 500\,kpc, and the generally successful fit to the concentration of galaxies at negative redshifts near 770\,kpc distance, largely the M\,31 satellites. At still greater distances most redshifts are positive but there is little correlation with distance. This illustrates the important influence of the mass outside the Local Group. 

\begin{table}[ht]
\centering
\begin{tabular}{lrrrrr}
\multicolumn{6}{c}{Table 6: Halo Shape Index}\\
\noalign{\medskip}
\tableline\tableline\noalign{\smallskip}
 $\alpha$ &    2.5 &    3.0 &    4.0 &    6.0 &   12.0 \\
 \tableline
$\chi^2$ &    459 &    404 &    368 &    338 &    333 \\
 $\langle N_\sigma^2\rangle$\ $cz$ &    1.8 &    1.3 &    1.1 &    0.8 &    0.7 \\
 $\langle N_\sigma^2\rangle$\ $d$ &    2.7 &    2.4 &    2.2 &    2.2 &    2.2 \\
 $\langle N_\sigma^2\rangle$\ $M$ &    4.1 &    4.6 &    4.5 &    4.5 &    4.6 \\
 $\langle N_\sigma^2\rangle$\ $\mu$ &    4.3 &    4.0 &    3.2 &    2.8 &    2.8 \\
 $\langle N_\sigma^2\rangle$\ $v_i$ &    0.4 &    0.4 &    0.4 &    0.4 &    0.4 \\
MW mass$^{\rm a}$ &     24 &     23 &     27 &     28 &     28 \\
$v_c({\rm MW})^{\rm b}$ &    229 &    223 &    215 &    217 &    217 \\
$r_c({\rm MW})^{\rm c}$ &    199 &    201 &    258 &    258 & 262 \\
${M/L_K}^{\rm d}$ &     36 &     35 &     28 &     27 &     28 \\
\tableline
 \noalign{\smallskip}
\multicolumn{6}{l}{$^{\rm a}$unit = $10^{11}M_\odot$ \ $^{\rm b}$km s$^{-1}$
\ $^{\rm c}$kpc\  $^{\rm d}$Solar}\label{fig.haloshape}
\end{tabular}
\end{table}

\subsection{Halo Shape Index}
Table 6 illustrates how the fit of model to catalog parameters depends on the halo shape index $\alpha$ in Equation~(\ref{eq:halomodel}). The first row is the $\chi^2$ sum over all parameters (Eq.~[\ref{eq:chisquared}]). By this measure the  overall fit is best for $\alpha = 6$ and 12 (the top two curves in Fig.~1).  These large values of $\alpha$ argue for a rather sharp transition between the inner region where the gravitational acceleration is proportional to the inverse first power of the radius and the outer inverse square law. I largely concentrate on results for the $\alpha=6$ model rather than $\alpha=12$, because the difference from the 
NFW halo shape is a little less pronounced. This is discussed in Section~\ref{sec:haloshape}. 

The next five rows in Table 6 show the mean square number of standard deviations of model from catalog for the indicated quantities. I attribute the fluctuations around the mean trends with $\alpha$ to the sensitivity of tracer orbits to slight changes of the complex interaction among the massive actors. The mean square number of standard deviations of model from catalog, $\langle N_\sigma^2\rangle$, for redshifts is close to unity. This is at least in part a result of the assigned lower bound on the redshift standard deviation, 5\,km\,s$^{-1}$. The assignment of a bound on the redshift standard deviation larger than the precision of the measurement of what is a subdominant part of each galaxy seems physically reasonable, but the result that the value of $\langle N_\sigma^2\rangle$ for redshifts is close to unity has limited significance. 

The mean $\langle N_\sigma^2\rangle$ for distances, in the third row, is high for all choices of $\alpha$. Again, this is in part an artifact of the choice of lower bound on the uncertainties in distance. But the large value of $\langle N_\sigma^2\rangle$ for distances is dominated by some seven tracer galaxies (Sec.\ref{sec:chisquared}), which argues for systematic errors in the model and perhaps in a few of the measurements. 

The value of $\langle N_\sigma^2\rangle$ for masses, in the fourth row, also is large for all $\alpha$. In particular, four of the twelve actor masses differ from catalog by three standard deviations, a factor of about two (Table~1).  I do not know any significant evidence that conflicts with this indication that the ratio of K-band luminosity to dark matter mass of a galaxy may scatter by a factor of two. 

The mean of $\langle N_\sigma^2\rangle$ for proper motions is lowest at  $\alpha\ga 6$, but still significantly larger than ideal. I do not know how to apportion this to systematic errors in the model and in the challenging proper motion measurements.

The computation assigned nominal standard deviation $\sigma=100$\,km\,s$^{-1}$ for the physical peculiar velocities  $v_i$ relative to the general expansion of the universe  at $1+z=10$. Few of the $v_i$ listed in Tables~2 and~3 are larger than this, and $\langle N_\sigma^2\rangle$ for the initial velocities is unrealistically small for all $\alpha$. That is, although the penalty on large $v_i$ plays an important role in the discovery of acceptable orbits, it seems that orbits with fairly reasonable fits to redshifts and distances tend to have smaller $v_i$ than I had anticipated. 

The MW circular velocity $v_c$ and the mean mass-to-light ratio $M/L_K$ do not contribute to $\chi^2$. Their values that minimize  $\chi^2$ given $\alpha$, and the effective MW radius (Eq.~[\ref{eq:halomodel}]), are listed in the last three rows of Table~6. 

 \begin{figure}[htpb]
\begin{center}
\includegraphics[angle=0,width=4.in]{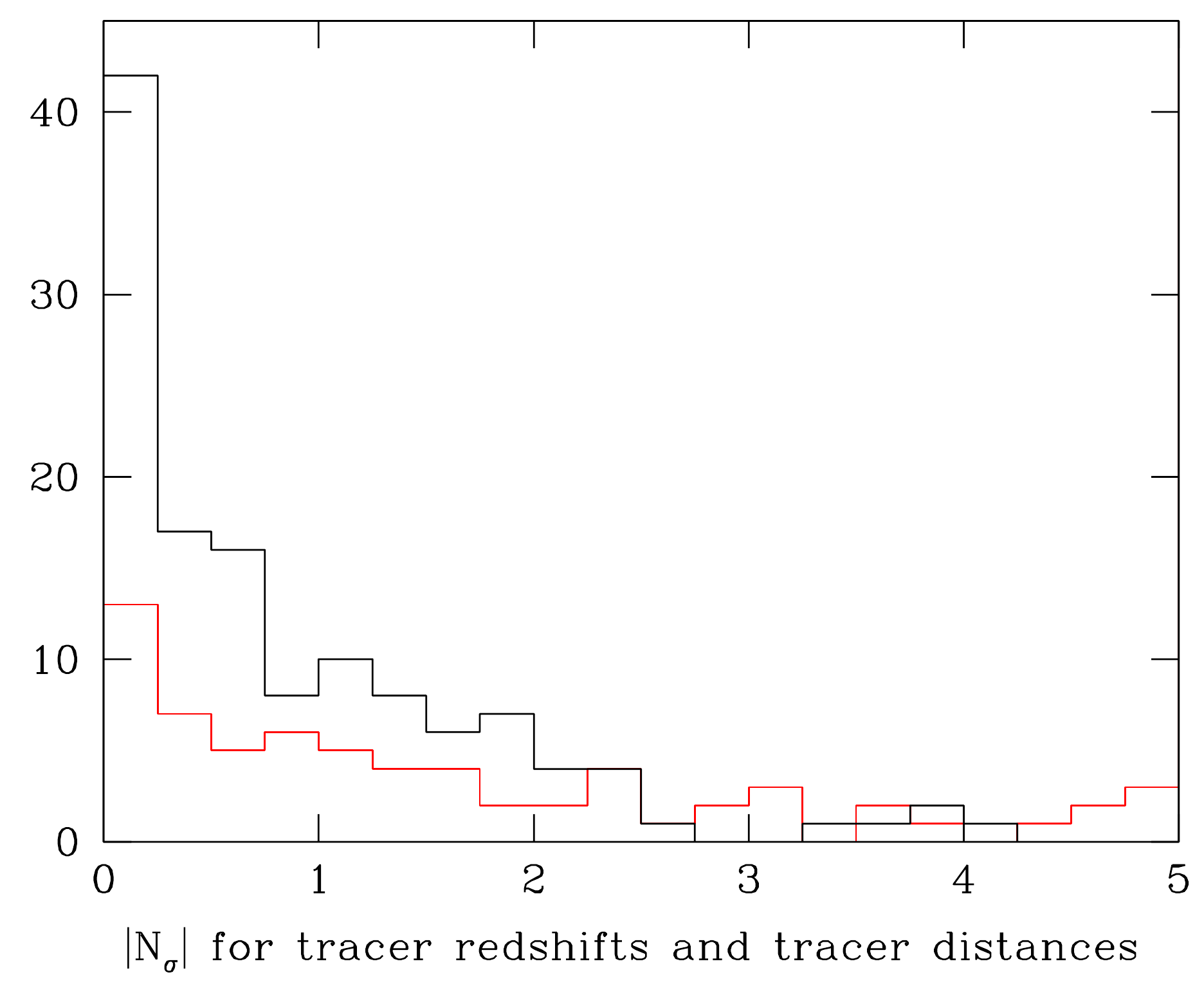} \vspace{-3mm}
\caption{\small Absolute values of numbers of standard deviations of differences of model and catalog values for the 64  tracer redshifts and 64 distances. The black histogram shows the results from $\alpha=6$ model, the red histogram the results from randomly assigned tracer angular positions.}\label{Fig:hist}
\end{center}
\end{figure}

\subsection{Rate of Accidental Fits to Redshifts and Distances}\label{sec:accidentals}

The mixed boundary conditions allow solutions to the equations of motion at each extremum or saddle point of the action. This means there can be multiple solutions among  which one naturally chooses the one with the lowest $\chi^2$. This can yield orbits that are wrong but accidentally offer reasonable-looking fits to the data.  I have checked the rate of accidentals by fitting the model to the data when the tracers are moved to random angular positions uniformly distributed over the sphere. The tracer redshifts are corrected for the change in contribution to the redshift by the component of the Solar velocity relative to MW along the line of sight. The catalog tracer distances are unchanged, and the orbits of the twelve massive actors are the same as in the $\alpha=6$ solution. If the tracers were moving in the close to spherically symmetric gravitational field of MW alone then the success rate for acceptable orbits would be the same for catalog and random angular positions. But since the other massive actors strongly break spherical symmetry an acceptable fit for a tracer with the wrong angular position likely is accidental. 

Figure~\ref{Fig:hist} shows distributions of the 128 $| N_\sigma |$ (Eq.~[\ref{eq:chisquared}]) for tracer redshifts and distances, the black histogram for catalog angular positions, the red histogram for random  angular positions. No $| N_\sigma |$ is off scale in the real catalog, while 61, about half the total, are off scale in the random catalog. But the red histogram does have a peak at small $| N_\sigma |$. It may be significant that the six closest tracers are more likely to have reasonably close fits to the measurements. These nearest tracers see the closest approximation to a spherically symmetric gravitational potential, making them least sensitive to a change of angular position. But it seems clear that the greater contribution to the peak at small $N_\sigma$ in the red histogram is the effect of multiple solutions that produce accidently reasonable fits to a redshift or distance. The considerably lower peak at small $N_\sigma$ for the random catalog, with half the $| N_\sigma |$ larger than any for the real catalog, tells us that  accidentals are subdominant in the black histogram. And accidentals might be expected to be less common in the real catalog because it offers the opportunity for physically significant fits. I conclude that accidentals likely have reduced the $\chi^2$ measure of the model fit to the data, but that they have not seriously contributed to the general agreement of model and catalog distances and redshifts scattered across Figure~2.

\begin{table}[ht]
\centering
\begin{tabular}{lcrrrrr}
\multicolumn{7}{c}{Table 7: 3-$\sigma$ Deviations of Model from Data}\\
\noalign{\medskip}
\tableline\tableline\noalign{\smallskip}
 &   & \multicolumn{5}{c}{$N_\sigma$} \\
\cline{3 - 7}
\multicolumn{2}{c}{Halo Shape Index} & 2.5 & 3 & 4 & 6 & 12 \\
\tableline
  MW             & mass     &$   1.9 $&$   1.8 $&$   3.5 $&$   3.6 $&$   3.6$  \\
  Maff               & mass     &$  -3.8 $&$  -4.2 $&$  -3.2 $&$  -3.1 $&$  -3.0$  \\
  M81                & mass     &$   2.6 $&$   2.9 $&$   3.5 $&$   3.5 $&$   3.5$  \\
  LMC                & redshift &$  -5.3 $&$  -3.5 $&$  -2.0 $&$  -1.0 $&$  -0.3$  \\
  M33                & mu delta &$  -3.4 $&$  -3.1 $&$  -3.1 $&$  -3.1 $&$  -3.0$  \\
  I10                & mu delta &$  -4.1 $&$  -4.1 $&$  -2.3 $&$  -2.1 $&$  -2.1$  \\
  Sextans (I)        & distance &$   2.7 $&$   3.3 $&$   1.4 $&$   0.4 $&$  -0.2$  \\
  Leo T              & distance &$   3.6 $&$   2.9 $&$  -2.9 $&$  -2.3 $&$  -2.1$  \\
  NGC 6822           & redshift &$  -3.2 $&$  -2.3 $&$  -2.2 $&$  -1.3 $&$  -0.4$  \\
  Tucana             & distance &$  -2.1 $&$  -2.2 $&$  -3.7 $&$  -3.9 $&$  -4.1$  \\
  Pegasus dIrr       & redshift &$  -3.2 $&$  -1.8 $&$  -1.4 $&$  -1.2 $&$  -0.7$  \\
  Pegasus dIrr       & distance &$  -4.1 $&$  -1.2 $&$  -1.1 $&$  -0.5 $&$  -0.4$  \\
  UGC 4879           & distance &$  -4.3 $&$  -4.6 $&$  -3.9 $&$  -3.9 $&$  -3.9$  \\
  KKR 25             & redshift &$  -4.1 $&$  -3.8 $&$  -3.5 $&$  -3.4 $&$  -3.4$  \\
  UGC 9128           & distance &$   4.4 $&$   4.4 $&$   4.3 $&$   4.2 $&$   4.2$  \\
  KKH 98             & distance &$   3.2 $&$   3.2 $&$   2.1 $&$   2.3 $&$   2.3$  \\
  KKH 86             & distance &$   4.1 $&$   3.9 $&$   3.7 $&$   3.7 $&$   3.6$  \\
\tableline
\end{tabular}
\end{table}

\subsection{Largest Contributions to $\chi^2$}\label{sec:chisquared}

The $\chi^2$ measure of fit is the sum over the 248 redshifts, distances, proper motions, and masses. Allowing for the freedom to adjust parameters one might expect $\chi^2\sim 170$ if model and standard deviations were sufficiently accurate.  Since the model certainly is incomplete, and many of the measurement uncertainties that are treated as standard deviations are at best informed estimates, while many  have even weaker provenance,  it is no surprise that the model value, $\chi^2\sim 340$ at $\alpha \ga 6$, is significantly larger. The value of $\chi^2$ thus has no formal meaning. But it seems significant that the excess value of the $\chi^2$ sum is dominated by the relatively small number of galaxies in Table~7.

Table~7 lists the galaxies for which $N_\sigma$ for any parameter exceeds three nominal standard deviations for any choice of the shape index $\alpha$. As for Table~6, I attribute fluctuations around the trends of $N_\sigma$ with $\alpha$ to the complex interactions of orbits of actors and tracers. The model redshift of LMC is seriously low at $\alpha = 2.5$, but reasonable at $\alpha = 6$ or 12. The model redshift and distance of Pegasus\,dIrr are low at small $\alpha$, acceptable at large $\alpha$. The proper motion of IC\,10 in the direction of increasing declination shows a similar though more modest trend. The distance to Tucana trends the other way, low at large $\alpha$, acceptable at two standard deviations at  $\alpha=2.5$. The difference of model from catalog distance to Leo\,T changes sign with increasing $\alpha$, and is arguably acceptable at  slightly more than  standard deviations  at large $\alpha$. 

The model masses of four of the actors (LMC is just under the cut for Table~7) are at $3\sigma$ from catalog, a factor of two. My impression is that we do not have the evidence to judge whether this is reasonable. The other five largest discrepancies at large $\alpha$, at about four nominal standard deviations, are the model distances of Tucana, UGC\,4879, UGC\,9128 and KKH\,86,  and the model redshift of KKR\,25. These are among the more distant of the tracers, at 1.4 to 2.6\,Mpc, where they are more at hazard of the imperfections of the schematic mass model. Clarification of their situation would be aided by more dwarf galaxy distance and redshift measurements at 1 to 3\,Mpc distance, which in turn likely would require a more detailed mass model. 

The distribution of $N_\sigma$ departures of model from catalog is exceedingly non-Gaussian. Eight tracers from the original McC sample contribute more than half the $\chi^2$ sum. When they are removed by putting them in two associations, then at large $\alpha$ five tracers in Table~7 contribute about half the excess of $\chi^2$ over what one might have hoped for in an adequate model. And that leaves 59 tracers whose model distances and redshifts are reasonably close to catalog, as illustrated in Figure~2.
 
  \begin{figure}[ht]
\begin{center}
\includegraphics[angle=0,width=4.5in]{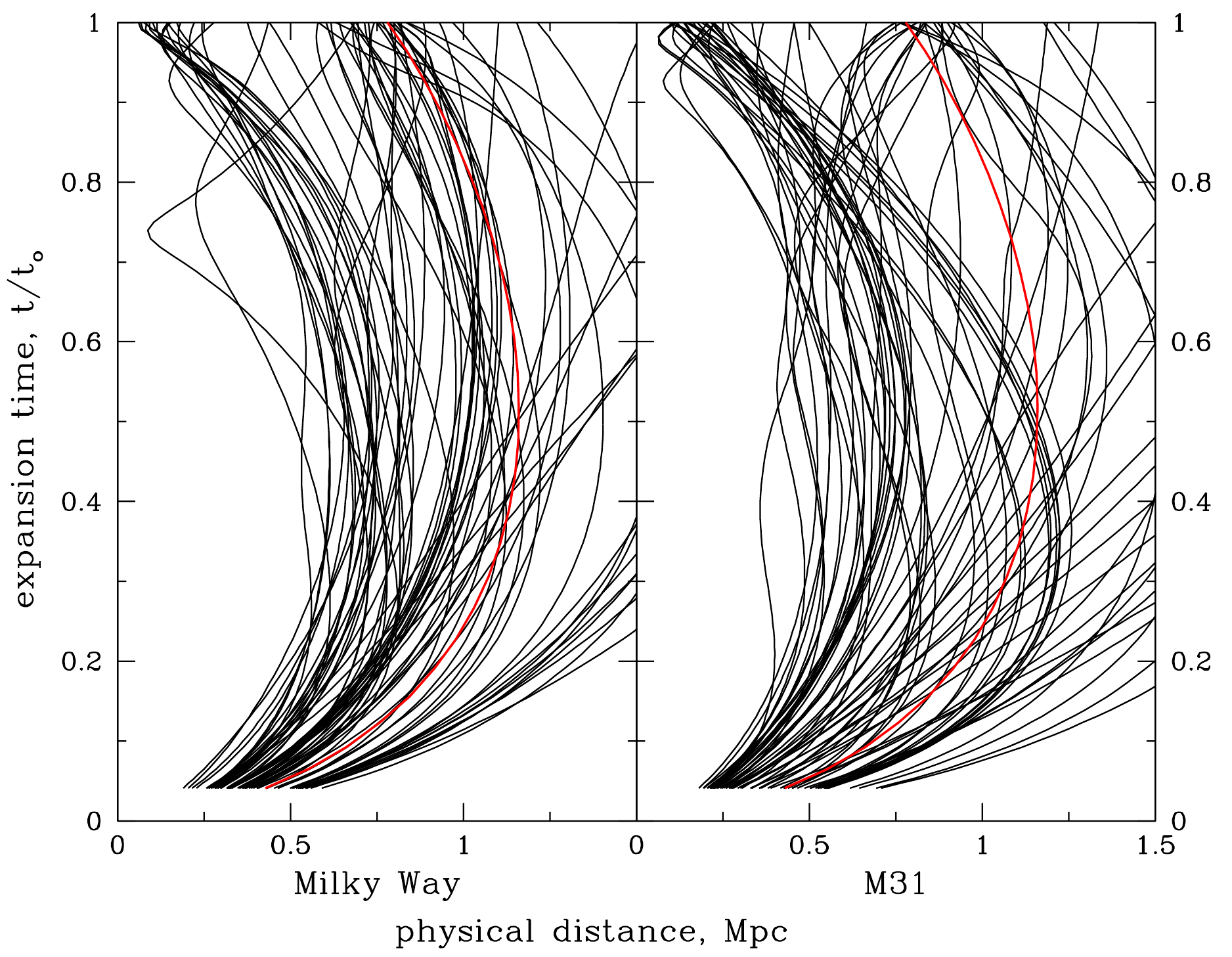} \vspace{-3mm}
\caption{\small Physical distances from MW as functions of physical time are shown in the left-hand panel, and physical distances from M\,31 in the right-hand panel, for $\alpha=6$.}\label{Fig:radial}
\end{center}
\end{figure}

\section{Orbits}\label{sec:orbits}

The left-hand panel in Figure~\ref{Fig:radial} shows the physical distances from MW as functions of physical time (relative to the present time) for all objects except the seven massive groups. The path of M\,31 is plotted in red. Some orbits converge toward MW, and others are seen to be in a stream that roughly follows M\,31. The right-hand panel shows the physical distances from M\,31 as functions of physical time, with MW plotted in red. These plots suggest that tracers now near MW  came from scattered initial positions, while the convergence toward M\,31 looks more stream-like. Caution is indicated, however, because the McC sample may be biased by the particularly deep survey for dwarfs close to M\,31 in the Pan-Andromeda Archaeological Survey\footnote{http://www.astro.uvic.ca/~alan/PANDAS/Home.html}.

 \begin{figure}[h]
\begin{center}
\includegraphics[angle=0,width=5.5in]{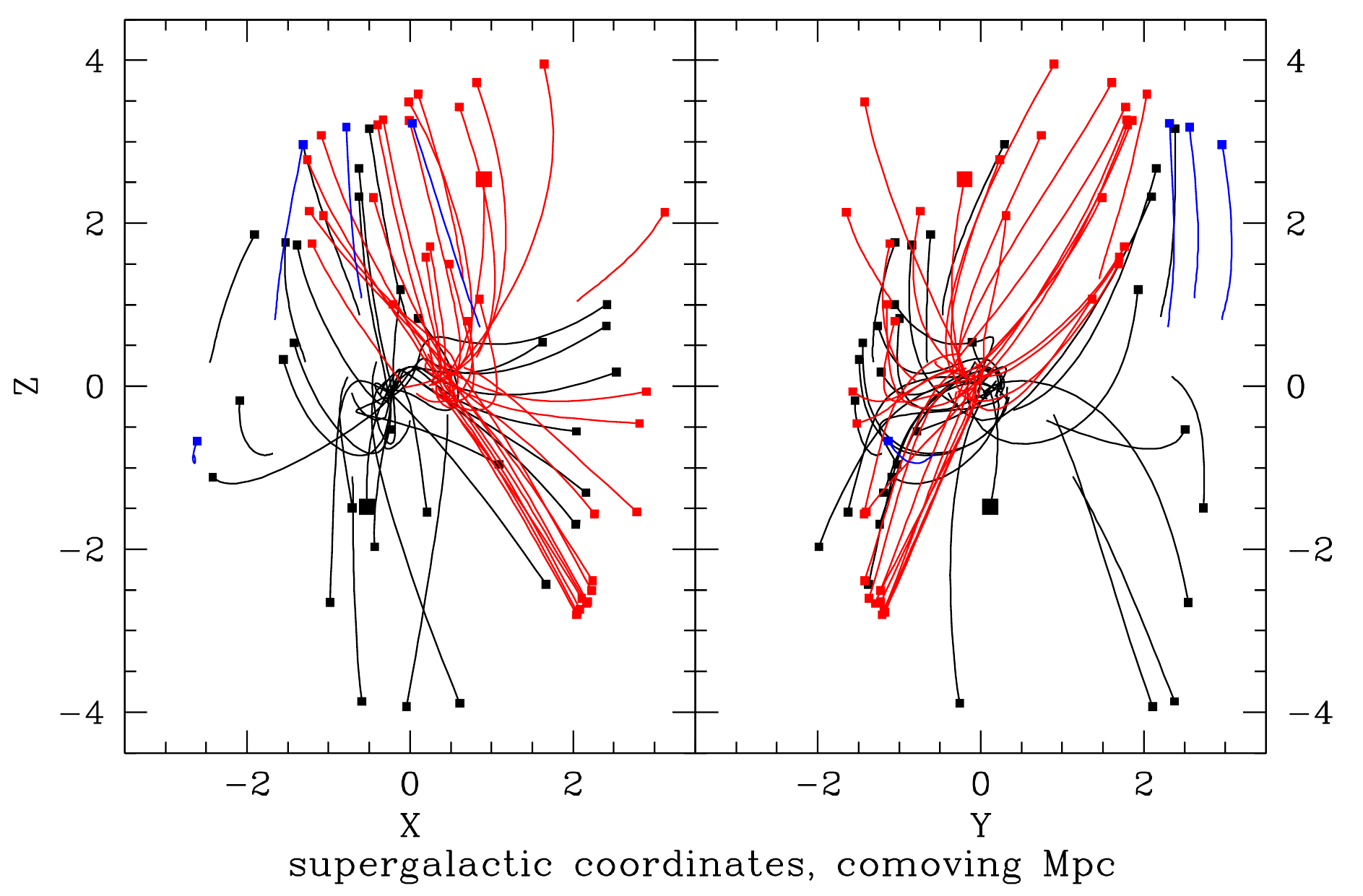} \vspace{-3mm}
\caption{\small Orbits of galaxies other than the massive groups in comoving supergalactic coordinates. Orbits that end up closest to MW are plotted in black, orbits now closest to M\,31 in red, and the other four in blue. Squares mark positions at $1+z=10$, the larger squares marking MW and M\,31.}\label{Fig:orbits}
\end{center}
\end{figure}
Figure~\ref{Fig:orbits} offers another way to compare the convergence of orbits toward the two galaxies. The model orbits of all but the seven massive groups are plotted in orthogonal projections in comoving supergalactic coordinates. The $Z$-axis is normal to the plane of the Local Supercluster. At each time step the origin of coordinates is the center of mass of MW and M\,31 at that time. The squares mark positions at the initial time in the solution, $1+z=10$, with larger squares for MW and M\,31. The squares and curves are black for the tracers now closest to MW, red for the tracers closest to M\,31. The remainder plotted in blue are UGC\,8508, now closest to M\,81; IC\,3104, closest to Cen; and UGC\,9128 and KKH\,86, closest to the M94 group. This figure follows the successive approximations to the behavior shown in Figure~3 in Peebles, Tully, and Shaya (2011), and in Figure~2 in Shaya and Tully (2013).

Figure~\ref{Fig:orbits} shows M\,31 approaching MW from above, roughly along the direction of the normal of the supergalactic plane. The initial positions of tracers now closest to MW seem to be little correlated with the orientation of the supergalactic plane. But some tracers now near M\,31 arrived in a tight stream from initial comoving distance $\sim 2.5$\,Mpc below the MW-M\,31 center of mass, at supergalactic longitude $\sim 135^\circ$. This tight stream does not seem likely to be an artifact of the deeper selection of dwarfs near M\,31. Most of the rest of the tracers now near M\,31 arrived in a broader cloud moving roughly with M\,31 from initial comoving positions  $\sim1.5$~to~4\,Mpc above the center of mass. Three of the blue orbits for tracers that are not now closest to MW or M\,31 originated above the plane alongside the concentration that ended up close to M\,31. The fourth blue orbit near the plane has not moved very far. 

 \begin{figure}[ht]
\begin{center}
\includegraphics[width=3.25in]{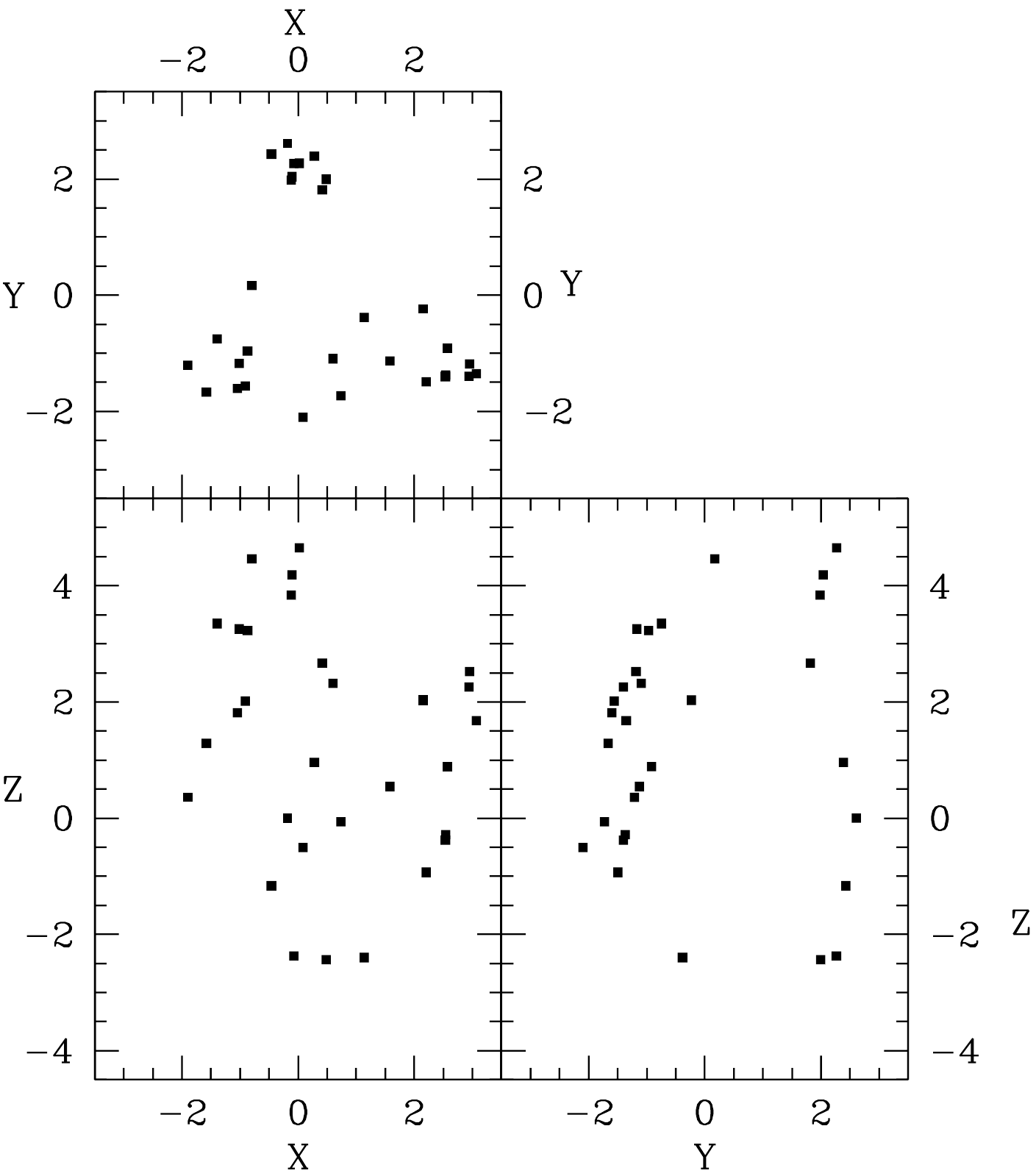}\includegraphics[width=3.25in]{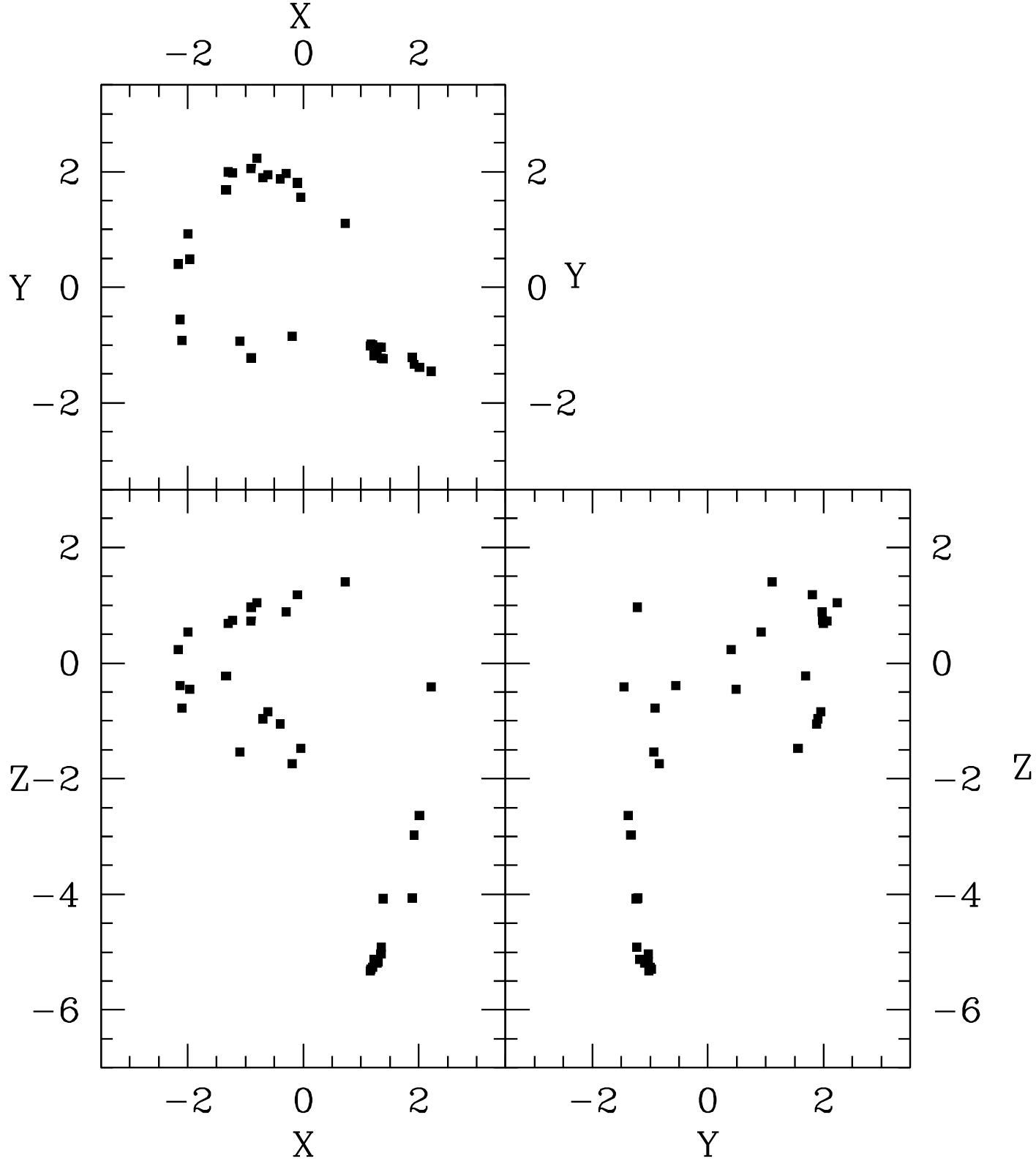}
\caption{\small Positions at $1+z=10$ of the tracers now closest to MW, at the origin of coordinates, in the left-hand panel, and those now closest to M\,31, at the origin of coordinates, in the right-hand panel.}\label{Fig:M31companions}
\end{center}
\end{figure}
The left-hand panel in Figure~\ref{Fig:M31companions} shows initial positions of the tracers now closest to MW, and the right-hand panel those now closest to M\,31. The orthogonal projections are plotted in supergalactic  coordinates, in units of comoving megaparsecs, with origin at the initial position of MW in the left-hand panel, and origin at M\,31 in the right-hand panel. The dwarfs now near these two galaxies are spread across regions of similar size at high redshift, but those now near MW have a clumpy initial pattern while those now near M\,31 approximate an interesting spiral pattern. Considerations of how the different characters of the orbits of dwarfs now near MW and M\,31 might relate to the differences of bulges of these two galaxies seem worth pursuing. But that might best await informed assessment of the effects of incompleteness caused by the zone of avoidance, the greater chance of discovering dwarfs closer to MW, and the particularly deep survey for dwarfs close to M\,31. Also awaiting analysis is the role of the mass distribution model, with its smooth massive groups, in determining positions of these tracer galaxies allowed by the condition of small initial peculiar velocities.

\section{Comparison to Other Measures of the Local Group}\label{sec:othermodels}

This computation allows the values of the K-band mass-to-light ratio and the MW circular velocity to float to minimize $\chi^2$, with the results in Table 6. Shaya and Tully~(2013) used $M/L_K = 40$ for spirals and 50 for ellipticals. Since there are only two $L\sim L_\ast$ ellipticals in the present study the Shaya and Tully value for spirals should be compared to the present result,  $M/L_K\sim 30$. In view of the considerable scatter of model from catalog masses (Table~1), the two values of $M/L_K$ seem reasonably consistent.  

The $\alpha=6$ solution puts the MW circular velocity at $v_c\simeq 217$~km\,s$^{-1}$. This is in the low velocity tail of the distribution of measurements compiled by Vall{\'e}e (2017). A complication is that the value of $v_c$ is most important in NAM for the translation between heliocentric quantities --- redshifts and proper motions --- and model values relative to the effective center of mass of the MW dark matter halo, which may be moving relative to the MW stars and gas. 

When the value of Hubble's constant is allowed to move to minimize $\chi^2$ it prefers $H_o\sim 80\hbox{ km s}^{-1}\hbox{ Mpc}^{-1}$. This is well above the assigned value in Equation~(\ref{eq:cosmology}) taken from the astronomical evidence, and even further from the evidence from the standard model for the CMB anisotropy. More data on nearby dwarf galaxies may help determine whether the model preference for larger $H_o$ is of any significance. 

 \begin{figure}[ht]
\begin{center}
\includegraphics[angle=0,width=4.5in]{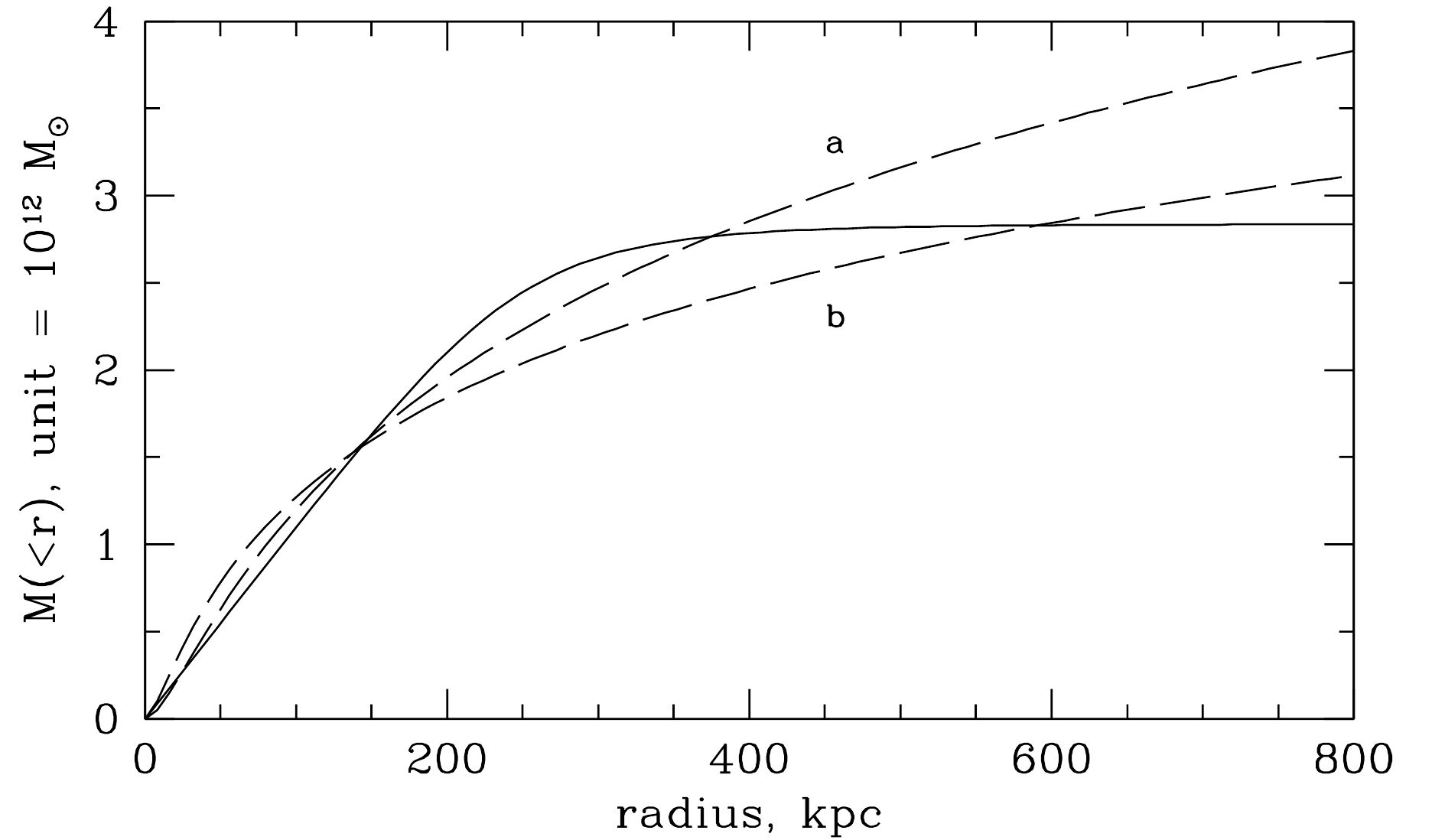} \vspace{-3mm}
\caption{\small Comparison of the  $\alpha = 6$ model for the MW mass within radius $r$ as a function of $r$ (the solid curve) to two NFW profiles (the dashed curves).\label{Fig:NFW}}
\end{center}
\end{figure}

\subsection{Shape of the Milky Way Halo}\label{sec:haloshape}

The $\chi^2$ measure of fit to the halo model in Equation~(\ref{eq:halomodel}) is minimized at halo shape index $\alpha\ga 6$ (Table~6). At $\alpha=6$ the MW halo mass $M(<r)$ within radius $r$ is plotted as the solid curve in Figure~\ref{Fig:NFW}. The break from the nearly flat inner rotation curve is sharper than in the two Navarro, Frenk, and White (1997) profiles plotted as dashed curves. In curve (a) the concentration parameter is $C=10$ and the characteristic radius is  $r_{200}=265$\,kpc; in curve (b), $C=20$ and $r_{200}=255$\,kpc. At radii $\la 400$\,kpc the two NFW profiles are not greatly from the $\alpha=6$ halo mass run, and might allow similar orbits for LMC and other closer galaxies. Of course, something must be done about the logarithmic divergence of the NFW halo mass at larger radius. The mass that would have been in the NFW profile (a) in MW and M\,31 at redshift $z=1$ and $r<800$\,kpc would have been rearranged by the gravitational interaction with the two galaxies and the massive groups. And fitting tracer orbits with this much mass  spread around the Local Group seems challenging.  One might postulate that profile (a) applies at $r\la 400$\,kpc and that the logarithmic tail was never present at larger radius. Curve (b) has a larger than conventional concentration parameter, which improves the situation at $r\sim800$\,kpc without greatly changing the mass within 50\,kpc. Here one might want to truncate the logarithmic tail at $r\sim 1$\,Mpc. With either truncation these NFW curves are not very far from the solid curve to which I have fitted the data. But the differences complicate comparisons of the present model results to measurements of the MW mass that assume the NFW profile.

\subsection{Masses of the Milky Way and the Local Group}\label{sec:masses}

Recent measures of the mass in the outer parts of MW based on fits to the NFW profile yielded NFW virial mass, respectively by Patel, Besla, and Mandel  (2017), McMillan (2017), and Fragione and Loeb (2017), 
\beqa
M_{\rm virial}&=&1.02^{+0.77}_{-0.75}\times 10^{12}M_\odot,\nonumber\\
&=& 1.30 \pm 0.30\times 10^{12}M_\odot,\\
&=& 1.2\hbox{ to }1.9\times 10^{12}M_\odot.\nonumber 
\eeqa
The consistency of these three results is encouraging. But comparison to the present NAM model result, 
\beq
M_{\rm total}= 2.8\times 10^{12}M_\odot,
\eeq
is complicated by the different  shapes of the MW halo models and the prescription needed to truncate the logarithmic divergence of the NFW profile. Within these uncertainties, I conclude that the total MW mass in the NAM model might be expected to be larger than the NFW virial mass, and that the excess of $M_{\rm total}$ over these estimates of $M_{\rm virial}$ is not obviously unreasonable. 

The Pe{\~n}arrubia, G{\'o}mez, Besla, et al. (2016) analysis of Local Group dynamics indicated Local Group mass  $2.64^{+0.42}_{-0.38}\times 10^{12} M_\odot$. The Local Group mass in the present solution is $4.5\times 10^{12} M_\odot$ (Table~1), some four standard deviations larger. The two analyses are based on similar redshifts,  distances, and proper motions. I suspect the difference is that the NAM model solution has larger transverse motions, which are more directly dynamically motivated. 

Eadie, Springford, and Harris (2017a,b) analyzed the galactic globular cluster positions in single-particle phase space, using measured positions and redshifts, proper motions when measured, and line-of-sight velocity components only when proper motions are not measured. Their MW halo mass within 125\,kpc is in the range 0.52 to $0.74 \times 10^{12}M_\odot$ at 95\% confidence. This is well below the mass $1.4\times 10^{12}M_\odot$ within 125\,kpc in the $\alpha=6$ solution. However, only 18 of the 143 globular clusters in the Eadie et al. analysis are at distances greater than 20\,kpc, so the derived mass within 20\,kpc might be expected to be more secure than the mass within 125\,kpc. The Eadie et al. sensitivity test indicates that when the phase space distribution is fitted to the 18 globulars more distant than 20\,kpc the derived mass within 125\,kpc is in the range 0.5 to  $1.75\times 10^{12}M_\odot$ at 95\% credibility. The mass in the $\alpha=6$ solution is in this range, but the potential remains for a serious inconsistency. Eadie et al. point out that their method of analysis may be applied to nearby dwarf galaxies, to distances within which the assumption of a time-independent distribution function in phase space seems reasonable. And the NAM analysis can be applied to the outermost galactic globular clusters whose orbits might be simple enough for NAM. It would be fascinating to see whether these approaches can bring the two measures of the mass within 125\,kpc to consistency.

In the NAM solution the MW mass is nearly twice that of M\,31.  This  mass ratio certainly is curious, but the NAM solution with these masses does offer an apparently reasonable fit to the measured redshifts and distances of most of the 64 dwarf galaxies, including the many near MW and M\,31. Of course, there is another curiosity in the behavior of the eight dwarfs in two association. Resolutions of these curiosities certainly call for more work. 

\subsection{Milky Way Escape Speed}

Williams, Belokurov, Casey, and Evans (2017) used the shape of the distribution of stellar radial velocities near the high velocity end to estimate the MW escape speed as a function of galactocentric distance. In the $\alpha=6$ solution I computed the escape speed to the maximum of the gravitational potential along the straight line connecting MW and M\,31, at 440\,kpc from MW. This should be quite close to the saddle point. The model escape speed at 8~kpc from the MW  is 600\,km\,s$^{-1}$. Williams et al. found local escape speed $521^{+46}_{-30}$\,km\,s$^{-1}$. The model escape speed at 50~kpc from the MW  is 425\,km\,s$^{-1}$. Williams et al. found escape speed $379^{+34}_{-28}$\,km\,s$^{-1}$ at this distance. These quantities seem to be reasonably  consistent. But caution is indicated because stars moving somewhat faster than escape speed might require longer than a Hubble time to find the saddle point toward M31, and high velocity stars in M\,31 may find their way across the saddle point to MW. That is, the highest velocity stars in the MW stellar halo could have larger galactocentric speeds than the escape speed to M\,31. 
 
\subsection{Galactocentric Velocities of M\,31 and the Nearby Dwarf Galaxies}

In the $\alpha=6$ solution the galactocentric velocity of M\,31 in galactic coordinates is
\beq
u = -63,\quad v = -168, \quad w = 38 \hbox{ km s}^{-1}, 
\eeq 
with radial and transverse components
\beq
V_{\rm radial} = -116,\quad V_{\rm tran} = 141\hbox{ km s}^{-1}.
\eeq
The van der Marel et al. (2012a) estimate of the transverse velocity is $V_{\rm tan} = 17\pm 34$ km s$^{-1}$. Although the model proper motion components of M\,31 differ from the  van der Marel et al. measurements by 2.3 and 1.0 standard deviations (Table 5), the transformation to galactocentric velocity introduces the  more serious-looking 3.6 standard deviation discrepancy in $V_{\rm tran}$. 

The mean galactocentric velocity of the 12 dwarf galaxies at model distances less than 150\,kpc from M31 is
\beq
u = 27,\quad v = -105, \quad w = 77 \hbox{ km s}^{-1}, \label{eq:stream1}
\eeq 
and the mean of the 11 dwarfs at model distances between 150 and 300\,kpc from M31 is
\beq
u = 37,\quad v = -64, \quad w = 30 \hbox{ km s}^{-1}.  \label{eq:stream2}
\eeq 
The rough similarity of mean velocities of these disjoint samples suggests the dwarfs near M\,31 have a meaningful streaming motion. However, the Salomon, Ibata, Famaey et al. (2016) analysis of the pattern of redshifts and angular positions of the dwarf galaxies around M31 indicated mean galactocentric streaming velocity (Salomon 2017)
\beq
u = 115,\quad v = 19, \quad w = 128 \hbox{ km s}^{-1}.  \label{eq:stream3}
\eeq
This differs from the model streaming velocities in Equations~(\ref{eq:stream1})  and~(\ref{eq:stream2}) by more than 100 km s$^{-1}$. 

A significant difference between the velocity of M\,31 and the mean motion of the dwarf galaxies near it does not seem unexpected: it is a signature of anisotropic flow of the dwarfs relative to M\,31, as in the tight stream approaching M\,31 from below the plane of the local supercluster (Fig.~\ref{Fig:orbits}). But the difference between independent estimates of the streaming flow of the dwarfs around this galaxy (in Eqs.~[\ref{eq:stream1}], [\ref{eq:stream2}], to be compared to Eq.~[\ref{eq:stream3}]) is a disturbing indication of our limited understanding of Local Group dynamics.

\subsection{Collisions}\label{sec:collisions}

There is a tradition of attributing low HI masses and peaks in the stellar age distributions in dwarf galaxies to disturbances by collisions with other dwarfs or close passages by massive galaxies. Serious collisions certainly happen, as in the Sagittarius Stream, and eventually the Magellanic Clouds. The galaxy orbits in this study do not show very close passages. This is at least in part because the analysis does not include dwarfs closer to MW than LMC, or galaxies with catalog positions closer than 100\,kpc from M\,31. The more isolated galaxies in this study are more likely to have avoided such effects. But collisions may be absent in part because NAM tends to avoid orbits with sharp accelerations. The best test for more interesting orbits that I may have missed may be a systematic comparison of orbits derived by NAM to orbits found by shooting back in time from conditions at low redshift similar to those from NAM, following Shaya and Tully (2013), as in their discussion of the issue of the NGC\,3109 association.

\subsection{Challenges to the Fits to Redshifts and Distances}\label{sec:challenges}

The Banik and Zhao (2017) analysis of Local Group dynamics yielded the ten most discrepant galaxies listed in their Table 4. We agree on serious problems with the four of their galaxies that I have lumped with two others in the NGC\,3109 association (Table~4). We also agree on the problem with UGC\,4879. It is in the list of most discrepant cases in Table~7, with model distance four standard deviations below catalog. The galaxies NGC\,55 and NCC\,4163 in the Banik and Zhao list are not treated as separate actors in this analysis: the former in the  Scl group, the latter in the  M94 group (Table~1). The motions of these galaxies may be complicated by proximity to these mass concentrations. The Banik and Zhao galaxy HIZSS\,3 is not in the McC catalog. Model and data for the last two, GR\,8 and KKR\,3, are not discrepant in my solutions (Table~3). The challenge of fitting redshifts and distances to all the nearby dwarf galaxies is real, but in my model largely confined to the two associations. Since it seems difficult to imagine the signature of new gravity physics considered by Banik and Zhao would be largely confined to the two associations, I turn to possible interpretations of the associations within standard physics. 

I used the NGC\,3109 and DDO\,210 associations to remove the eight galaxies listed in Table~4 that present the greatest discrepancies from the redshift and distance measurements when treated as massless tracer particles. Two aspects of these galaxies seem significant. First, they are in two compact ranges of position and redshift. One is prominent enough to be named; I have not found discussions of the other. Second, the model distances in the associations are systematically larger than catalog and the model redshifts are systematically smaller than catalog. Shaya and Tully (2013) argue that the redshifts and distances in the NGC\,3109 association may be the result of an early close interaction with MW. The idea certainly deserves further consideration. 

I have adopted the working assumption that each association is gravitationally bound. At radius $r\sim 300$\,kpc and internal velocity $v\sim 100$\,km\,s$^{-1}$ the mass of NGC\,3109 would be roughly $v^2r/G\sim 10^{12}M_\odot$. This is comparable to the MW mass, but with luminosity $L_K\sim 10^8L_\odot$ it requires $M/L_K\sim 10^4$. For DDO\,210, $r\sim 200$\,kpc, $v\sim 30$ \,km\,s$^{-1}$, $v^2r/G\sim 10^{11}M_\odot$, comparable to the LMC mass, but at $L_K\sim 10^6L_\odot$ it requires $M/L_K\sim 10^5$. These would be unprecedented examples of dark matter halos with masses comparable to ordinary galaxies but far less gas and stars. The idea could be explored by treating the associations as massive actors in an analysis that is otherwise the same as here. 

A less dramatic idea to be explored is that the associations are transient, concentrations of galaxies that are not gravitationally bound but happen to be passing each other at the present epoch, perhaps closely enough that is not a good approximation to ignore their masses. Analysis of how mass-to-light ratios more modest than in the gravitationally bound hypothesis might alter line of sight velocities in the two associations enough to fit the measurements seems to be feasible within the present methods. I hope to report on this idea in due course. 

\section{Concluding Remarks}\label{sec:concludingremarks}

This analysis might have been challenged by pure dark matter simulations of structure growth in the established $\Lambda$CDM cosmology, for they suggest galaxies grew by major mergers at redshifts $z \la 2$. But most of the  $L\sim L_\ast$ galaxies within 10\,Mpc have classical bulge-to-total luminosity ratios less than 0.1 (Kormendy, Drory, Bender, and Cornell 2010; Fisher and Drory 2011). These galaxies cannot have grown by serious mergers of stellar systems; they had to have grown by a gentle rain of diffuse gas or plasma. And under this condition one might expect that the NAM model orbits usefully approximate the motion of the center of mass of the matter locally coalescing into a galaxy. But the direct test is the ability of the model to fit measured redshifts and distances. Figure~\ref{Fig:dz} shows that the model does quite well. 

The model is challenged by the multiple orbits allowed by the mixed boundary conditions. The solution presented here likely includes some orbits of dwarf galaxies that are wrong but happen to fit the catalog redshifts and distances. The test with randomly assigned angular positions (Sec.~\ref{sec:accidentals}) shows this has not significantly biased the results, but a check with more nearby dwarfs would add welcome weight to the test. There may be other solutions I have not found that allow an equally good or even better fit to the data, or maybe allow the M\,31 mass to be closer to that of MW. Checking this by a more thorough exploration of more intelligently chosen trial orbits likely would require a more capably organized computation. 

This model and others are challenged on the observational side by apparent inconsistencies, clear evidence of our limited understanding of dynamics of the nearby galaxies. Within the NAM model the most interesting challenge is the behavior of the associations NGC\,3109 and DDO\,210 (Table~4). Five other galaxies listed in Table~7 have $\sim 4$-$\sigma$ discrepancies, far more than would be expected if model and data were sufficiently accurate. This situation might be improved by closer attention to a mass model that better represents the distribution of more luminous galaxies further than 1\,Mpc from the Local Group. Also to be contemplated is the possibility that some of these five anomalous cases reflect underestimated uncertainties in the distances to the more distant dwarfs in this sample. 

The addition of reliable redshifts and distances of even more dwarfs between 50\,kpc and 1\,Mpc could be readily analyzed within the present model, and would add most welcome weight to the test of the model. Additions to the data at distances 1 to 3\,Mpc would allow a most welcome extension of this analysis, and likely would merit and require a better mass model. And in a better model at these greater distances it might be appropriate to explore the effect of tides from the more distant mass distribution, as in Shaya and Tully (2013). I expect that for a larger sample to 3\,Mpc distance this would best be done using a free parameterized tidal field model. A positive detection of tides would allow an interesting test, a comparison to the tide expected from what is known about the large-scale distributions of galaxies and mass. 

Pending additions to the catalog of nearby dwarfs, the argument for the NAM approach presented here is that it agrees with the measured redshifts and distances of the considerable number of McC galaxies around the Milky Way and the Andromeda Nebula, as shown in the redshift-distance plot in Figure~\ref{Fig:dz}. 

\acknowledgments
I have profited from discussions with Gwen Eadie,  Bill Harris, Alan McConnachie, Jean-Baptiste Salomon, Ed Shaya, and Brent Tully.


\begin{thebibliography}{}

\bibitem[Banik \& Zhao(2016)]{2017MNRAS.467} Banik, I., \& Zhao, H.\ 2017, \mnras, 467, 2180 

\bibitem[Brunthaler et al.(2005)]{2005Sci...307.1440B} Brunthaler, A., Reid, M.~J., Falcke, H., Greenhill, L.~J., \& Henkel, C.\ 2005, Science, 307, 1440 

\bibitem[Brunthaler et al.(2007)]{2007A&A...462..101B} Brunthaler, A., Reid, M.~J., Falcke, H., Henkel, C., \& Menten, K.~M.\ 2007, \aap, 462, 101 

\bibitem[Casetti-Dinescu \& Girard(2016)]{2016MNRAS.461..271C} Casetti-Dinescu, D.~I., \& Girard, T.~M.\ 2016, \mnras, 461, 271 

\bibitem[Corbelli et al.(2010)]{2010A&A...511A..89C} Corbelli, E., Lorenzoni, S., Walterbos, R., Braun, R., \& Thilker, D.\ 2010, \aap, 511, A89 

 
\bibitem[Eadie et al.(2017)]{2017ApJ...835..167E} Eadie, G.~M., Springford, A., \& Harris, W.~E.\ 2017a, \apj, 835, 167 

\bibitem[Eadie et al.(2017)]{2017ApJ...838...76E} Eadie, G.~M., Springford, A., \& Harris, W.~E.\ 2017b, \apj, 838, 76 

\bibitem[Fisher \& Drory(2011)]{2011ApJ...733L..47F} Fisher, D.~B., \& Drory, N.\ 2011, \apjl, 733, L47 

\bibitem[Fragione \& Loeb(2017)]{2017NewA...55...32F} Fragione, G., \& Loeb, A.\ 2017, \na, 55, 32 

\bibitem[Freedman et al.(2012)]{2012ApJ...758...24F} Freedman, W.~L., Madore, B.~F., Scowcroft, V., et al.\ 2012, \apj, 758, 24 

\bibitem[Kallivayalil et al.(2013)]{2013ApJ...764..161K} Kallivayalil, N., van der Marel, R.~P., Besla, G., Anderson, J., \& Alcock, C.\ 2013, \apj, 764, 161 

\bibitem[Kormendy et al.(2010)]{2010ApJ...723...54K} Kormendy, J., Drory, N., Bender, R., \& Cornell, M.~E.\ 2010, \apj, 723, 54 

\bibitem[McConnachie(2012)]{2012AJ....144....4M} McConnachie, A.~W.\ 2012, \aj, 144, 4 

\bibitem[McConnachie(2015)]{2015} McConnachie, A.~W.\ 2015, \url{www.astro.uvic.ca/~alan/Nearby_Dwarf_Database.html}

\bibitem[McMillan(2017)]{2017MNRAS.465...76M} McMillan, P.~J.\ 2017, \mnras, 465, 76 

\bibitem[McQuinn et al.(2015)]{2015ApJ...812..158M} McQuinn, K.~B.~W., Skillman, E.~D., Dolphin, A., et al.\ 2015, \apj, 812, 158 

\bibitem[Navarro et al.(1997)]{1997ApJ...490..493N} Navarro, J.~F., Frenk, C.~S., \& White, S.~D.~M.\ 1997, \apj, 490, 493 

\bibitem[Patel et al.(2017)]{2017arXiv170305767P} Patel, E., Besla, G., \& Mandel, K.\ 2017, arXiv:1703.05767  %

\bibitem[Pawlowski \& McGaugh(2014)]{2014MNRAS.440..908P} Pawlowski, M.~S., \& McGaugh, S.~S.\ 2014, \mnras, 440, 908 

\bibitem[Peebles(2009)]{2009arXiv0907.5207P} Peebles, P.~J.~E.\ 2009, arXiv:0907.5207 

\bibitem[Peebles(2010)]{2010arXiv1009.0496P} Peebles, P.~J.~E.\ 2010, arXiv:1009.0496 

\bibitem[Peebles et al.(2011)]{2011arXiv1105.5596P} Peebles, P.~J.~E., Tully, R.~B., \& Shaya, E.~J.\ 2011, arXiv:1105.5596 

\bibitem[Peebles \& Tully(2013)]{2013arXiv1302.6982P} Peebles, P.~J.~E., \& Tully, R.~B.\ 2013, arXiv:1302.6982 

\bibitem[Pe{\~n}arrubia et al.(2016)]{2016MNRAS.456L..54P} Pe{\~n}arrubia, J., G{\'o}mez, F.~A., Besla, G., Erkal, D., \& Ma, Y.-Z.\ 2016, \mnras, 456, L54 

\bibitem[Pe{\~n}arrubia et al.(2014)]{2014MNRAS.443.2204P} Pe{\~n}arrubia, J., Ma, Y.-Z., Walker, M.~G., \& McConnachie, A.\ 2014, \mnras, 443, 2204 

\bibitem[Riess et al.(2016)]{2016ApJ...826...56R} Riess, A.~G., Macri, L.~M., Hoffmann, S.~L., et al.\ 2016, \apj, 826, 56 

\bibitem[Salomon et al.(2016)]{2016MNRAS.456.4432S} Salomon, J.-B., Ibata, R.~A., Famaey, B., Martin, N.~F., \& Lewis, G.~F.\ 2016, \mnras, 456, 4432 

\bibitem[Salomon(2017)]{,,,} Salomon, J.-B. 2017, private communication 

\bibitem[Sand et al.(2015)]{2015ApJ...812L..13S} Sand, D.~J., Spekkens, K., Crnojevi{\'c}, D., et al.\ 2015, \apjl, 812, L13 

\bibitem[Sch{\"o}nrich et al.(2010)]{2010MNRAS.403.1829S} Sch{\"o}nrich, R., Binney, J., \& Dehnen, W.\ 2010, \mnras, 403, 1829 

\bibitem[Shaya \& Tully(2013)]{2013MNRAS.436.2096S} Shaya, E.~J., \& Tully, R.~B.\ 2013, \mnras, 436, 2096 

\bibitem[Sohn et al.(2013)]{2013ApJ...768..139S} Sohn, S.~T., Besla, G., van der Marel, R.~P., et al.\ 2013, \apj, 768, 139 

\bibitem[Tully(2014)]{2014privatecommunication} Tully, R. B. 2014, private communication

\bibitem[van der Marel et al.(2012a)]{2012ApJ...753....8V} van der Marel, R.~P., Fardal, M., Besla, G., et al.\ 2012, \apj, 753, 8

\bibitem[Vall{\'e}e(2017)]{2017Ap&SS.362...79V} Vall{\'e}e, J.~P.\ 2017, \apss, 362, 79 

\bibitem[van der Marel et al.(2012b)]{2012ApJ...753....9V} van der Marel, R.~P., Besla, G., Cox, T.~J., Sohn, S.~T., \& Anderson, J.\ 2012, \apj, 753, 9

\bibitem[van der Marel \& Sahlmann(2016)]{2016ApJ...832L..23V} van der Marel, R.~P., \& Sahlmann, J.\ 2016, \apjl, 832, L23 

\bibitem[Williams et al.(2017)]{2017MNRAS.468.2359W} Williams, A.~A., Belokurov, V., Casey, A.~R., \& Evans, N.~W.\ 2017, \mnras, 468, 2359 

\end{thebibliography}
\end{document}